\documentclass[11pt,preprintnumbers,titlepage,nofootinbib]{revtex4-2}
\usepackage[latin9]{inputenc}
\usepackage{array}
\usepackage{float}
\usepackage{multirow}
\usepackage{amsmath}
\usepackage{amssymb}
\usepackage{graphicx}
\usepackage[unicode=true,
 bookmarks=false,
 breaklinks=false,pdfborder={0 0 1},backref=section,colorlinks=false]
 {hyperref}
\hypersetup{
 colorlinks,linkcolor=blue,citecolor=blue,urlcolor=blue}

\makeatletter

\providecommand{\tabularnewline}{\\}
\newcommand{\lyxdot}{.}

\usepackage{wasysym}
\usepackage{array}
\usepackage{multirow}
\usepackage{wasysym}
\usepackage{wasysym}
\usepackage{amsfonts}
\usepackage{subfigure}
\usepackage{cleveref}
\usepackage{listings}
\usepackage{xcolor}
\usepackage{array}
\usepackage{url}
\usepackage{color}
\usepackage{subfigure}

\providecommand{\tabularnewline}{\\}
\@ifundefined{textcolor}{}{
\definecolor{BLACK}{gray}{0}
\definecolor{WHITE}{gray}{1}
\definecolor{RED}{rgb}{1,0,0}
\definecolor{GREEN}{rgb}{0,1,0}
\definecolor{BLUE}{rgb}{0,0,1}
\definecolor{CYAN}{cmyk}{1,0,0,0}
\definecolor{MAGENTA}{cmyk}{0,1,0,0}
\definecolor{YELLOW}{cmyk}{0,0,1,0}
\definecolor{ballblue}{rgb}{0.13, 0.67, 0.8}
\definecolor{bleudefrance}{rgb}{0.19, 0.55, 0.91}
\definecolor{blue(ncs)}{rgb}{0.0, 0.53, 0.74}
\definecolor{darkpastelgreen}{rgb}{0.01, 0.75, 0.24}
\definecolor{darkspringgreen}{rgb}{0.09, 0.45, 0.27}
\definecolor{denim}{rgb}{0.08, 0.38, 0.74}
\definecolor{electricviolet}{rgb}{0.56, 0.0, 1.0}
}

\makeatother

\begin{document}
\preprint{CTP-SCU/2024001}
\title{Distinguishing the Observational Signatures of Hot Spots Orbiting
Reissner-Nordström Spacetime}
\author{Tianshu Wu}
\email{wutianshu@stu.scu.edu.cn}

\author{Yiqian Chen}
\email{yqchen@stu.scu.edu.cn}

\affiliation{Center for Theoretical Physics, College of Physics, Sichuan University,
Chengdu, 610064, China}
\begin{abstract}
This paper delves into observable signatures of hot spots orbiting
Reissner-Nordström (RN) black holes and naked singularities. In a
RN black hole case, we find two discernible lensing image tracks in
time integrated images capturing a complete orbit of hot spots, and
a image shadow within the critical curve where photons with a small
impact parameter fall into the event horizon. Conversely, in RN singularities,
additional image tracks can be found inner the critical curve, originating
from photons reflected by the infinitely high effective potential
well. Moreover, we found incomplete and converge tracks from the time
integrated images of hot spot orbiting RN singularities lacking of
a photon sphere. The presence of these additional image tracks exerts
a significant influence on temporal magnitudes at their local maxima,
allowing us to differentiate between RN black holes and RN naked singularities.
\end{abstract}
\maketitle
\tableofcontents{}

\section{Introduction}

\label{sec:Introduction}

Recent advancements in gravitational physics have been propelled by
a surge in high-precision experiments, notably spearheaded by collaborations
like the Event Horizon Telescope (EHT). They have yielded groundbreaking
discoveries, including the detection of imaging of shadow-like features
near supermassive black holes \citep{Akiyama:2019cqa,Akiyama:2019brx,Akiyama:2019sww,Akiyama:2019bqs,Akiyama:2019fyp,Akiyama:2019eap,Akiyama:2021qum,Akiyama:2021tfw,EventHorizonTelescope:2022xnr,EventHorizonTelescope:2022vjs,EventHorizonTelescope:2022wok,EventHorizonTelescope:2022exc,EventHorizonTelescope:2022urf,EventHorizonTelescope:2022xqj}.
These experimental triumphs align closely with theoretical predictions,
notably supporting the Kerr hypothesis \citep{Will:2014kxa}, which
posits that the endpoint of gravitational collapse in suitable astrophysical
conditions results in the formation of rotating, electrically neutral
black holes \citep{Kerr:1963ud,Penrose:1964wq}. This convergence
of theory and observation not only validates our understanding of
black hole physics but also opens avenues for exploring unresolved
questions in the realm of strong gravitational fields. Furthermore,
physicists continue to refine simplified models to numerically explore
black hole properties, including accretion disks and celestial spheres,
enhancing our understanding of these enigmatic cosmic entities.

A defining feature of images captured by the EHT is a dark, central
region surrounded by a bright ring. This feature emerges from the
bending of light due to strong gravity near unstable photon orbits
\citep{Synge:1966okc,Bardeen:1972fi,Bardeen:1973tla,Virbhadra:1999nm,Claudel:2000yi,Virbhadra:2008ws,Bozza:2009yw,Virbhadra:2022iiy}.
These observations align well with the Kerr black hole model. However,
the discovery of photon spheres around exotic compact objects (ECOs)
has introduced complexities \citep{Schmidt:2008hc,Guzik:2009cm,Liao:2015uzb,Goulart:2017iko,Nascimento:2020ime,Islam:2021ful,Tsukamoto:2021caq,Junior:2021svb,Olmo:2021piq,Ghosh:2022mka}.
ECOs mimic black hole observations but predict unique signatures for
further verification. Therefore, their observation becomes a scientific
target for the next-generation Event Horizon Telescope (ngEHT) \citep{Ayzenberg:2023hfw}.

Naked singularities, a type of ECO, have garnered significant interest.
Despite the cosmic censorship conjecture forbidding naked singularities,
they can theoretically form under specific conditions during the collapse
of massive objects \citep{Shapiro:1991zza,Joshi:1993zg,Harada:1998cq,Joshi:2001xi,Goswami:2006ph,Banerjee:2017njk,Bhattacharya:2017chr}.
With photon spheres, naked singularities can closely resemble black
holes optically, prompting investigations into their unique observational
signatures \citep{Virbhadra:2002ju,Virbhadra:2007kw,Gyulchev:2008ff,Sahu:2012er,Shaikh:2019itn,Paul:2020ufc,Zhdanov:2019ozq,Stashko:2021lad,Stashko:2021het,Tsukamoto:2021fsz,Wang:2023jop,Chen:2023trn}.
Notably, research indicates that naked singularities in Reissner-Nordström
(RN) geometries possess an anti-photon sphere, reversing the inner
image that does not match the observed shadow \citep{Shaikh:2019itn}.
Therefore, investigating the distinct observational characteristics
of RN naked singularities and RN black holes holds significant value.

On the other hand, several general relativistic magnetohydrodynamics
(GRMHD) simulations and semi-analytic models suggest that the interplay
of magnetic reconnection and flux eruptions can lead to the formation
of hot spots around supermassive black holes residing within magnetized
accretion disks \citep{Dexter:2020cuv,Scepi:2021xgs,ElMellah:2021tjo}.
Interestingly, these hot spots have been repeatedly observed in the
vicinity of Sgr A{*} \citep{Witzel:2020yrp,Michail:2021pgd,GRAVITY:2021hxs}.
Moreover, an intriguing instance involved the detection of an orbiting
hot spot in unresolved light curve data at the EHT observing frequency
\citep{Wielgus:2022heh}. Due to their origin from a compact region
very close to the innermost stable circular orbit (ISCO), these hot
spots offer a promising avenue for probing central objects in the
strong gravity regime \citep{Hamaus:2008yw,abuter2018detection,Rosa:2022toh,Rosa:2023qcv,Tamm:2023wvn,Rosa:2024bqv}.
This work focuses on investigating the distinctive signatures of hot
spots around different RN singularities and RN black holes to potentially
distinguish them.

The subsequent sections of this paper are structured as follows: In
Section \ref{sec:Set up}, we briefly introduce RN spacetime, along
with a discussion of geodesic motion within these spacetimes. In Section
\ref{sec:Celestial-Sphere-Model}, we show the numerical results of
the celestial model in RN spacetime. Section \ref{sec:Hot Spot Model}
is devoted to the hot spot model, followed by an examination time
integrated images, temporal magnitudes and centroids. Finally, Section
\ref{sec:CONCLUSIONS} presents our conclusions. We adopt the convention
$G=c=1$ throughout the paper.

\section{Set up}

\label{sec:Set up}

This section commences with a concise overview of geodesic expressions
within RN spacetime. To enhance the significance of the chosen orbit
in our hot spot model, we also provide a brief introduction to diverse
stable circular orbits within the framework of RN spacetime.

The RN metric presents a static solution in Einstein and Maxwell equations,
which is given by 
\begin{equation}
ds^{2}=-f\left(r\right)dt^{2}+\frac{1}{f\left(r\right)}dr^{2}+r^{2}\left(d\theta^{2}+\sin^{2}\theta d\varphi^{2}\right),\label{eq:metric-1-1}
\end{equation}
with the metric function 
\begin{equation}
f\left(r\right)=1-\frac{2M}{r}+\frac{Q^{2}}{r^{2}}.
\end{equation}
Here, $M$ and $Q$ represent the mass and the electric charge, respectively.
Conditions $M<Q$ and $M>Q$ correspond to the two scenarios of naked
singularities and black holes, respectively. The trajectory of a test
particle with four-momentum $p^{\mu}$ is determined by the geodesic
equations
\begin{equation}
\frac{dx^{\mu}}{d\lambda}=p^{\mu},\quad\frac{dp^{\mu}}{d\lambda}=-\Gamma_{\rho\sigma}^{\mu}p^{\rho}p^{\sigma},\label{eq:geoeq-1}
\end{equation}
where $\lambda$ is the affine parameter, and $\Gamma_{\rho\sigma}^{\mu}$
indicates the Christoffel symbol. These geodesics are fully characterized
by three conserved quantities, 
\begin{equation}
E=-p_{t},\quad L_{z}=p_{\varphi},\quad L^{2}=p_{\theta}^{2}+L_{z}^{2}\csc^{2}\theta.
\end{equation}

In the context of massless particles, the conserved quantities $E$,
$L_{z}$ and $L$ represent the total energy, the angular momentum
parallel to the axis of symmetry and the total angular momentum, respectively.
Additionally, the Hamiltonian constraint $\mathcal{H}\equiv g_{\mu\nu}p^{\mu}p^{\nu}/2=0$
yields the radial component of the null geodesic equations as
\begin{equation}
\dot{r}^{2}+V_{\text{eff}}\left(r\right)=0,
\end{equation}
where the dot signifies differentiation with respect to an affine
parameter $\lambda$, and the introduced effective potential is given
by 
\begin{equation}
V_{\text{eff}}\left(r\right)=f\left(r\right)\left[\frac{L^{2}}{r^{2}}-\frac{E^{2}}{f\left(r\right)}\right].
\end{equation}

A circular null geodesic occurs at an extremum of the effective potential
$V_{\text{eff}}(r)$, and the radius $r_{c}$ of this geodesic is
determined by the conditions 
\begin{equation}
V_{\text{eff}}\left(r_{c}\right)=0,\text{ }V_{\text{eff}}^{\prime}\left(r_{c}\right)=0.
\end{equation}
Furthermore, local maxima and minima of the effective potential correspond
to unstable and stable circular null geodesics, which constitute the
photon sphere and the anti-photon sphere, respectively. When $Q/M\leq\sqrt{9/8}$,
a photon sphere exists at $r=r_{ps}^{+}$, and an anti-photon sphere
exists at $r=r_{ps}^{-},$ as given by
\begin{equation}
r_{ps}^{\pm}=\frac{3M\pm\sqrt{9M^{2}-8Q^{2}}}{2}.
\end{equation}

For massive particles, $E$, $L_{z}$ and $L$ represent the total
energy per unit mass, the angular momentum per unit mass parallel
to the axis of symmetry and the total angular momentum per unit mass,
respectively, when the affine parameter $\lambda$ is chosen as the
proper time per unit mass. Similarly, the Hamiltonian constraint $\mathcal{H}=-1/2$
leads to the effective potential 
\begin{equation}
V_{\text{eff}}\left(r\right)=f\left(r\right)\left[\frac{L^{2}}{r^{2}}-\frac{E^{2}}{f\left(r\right)}+1\right].
\end{equation}
and circular time-like geodesics satisfy 
\begin{equation}
V_{\text{eff}}\left(r\right)=0\text{, }V_{\text{eff}}^{\prime}\left(r\right)=0.\label{eq:MSCO1}
\end{equation}

A comprehensive analysis of the circular motion of time-like particles
have been studied in \citep{Pugliese:2010ps,Pugliese:2011py}. In
the presence of a photon sphere $(Q/M\leq\sqrt{9/8})$ and ensuring
the motion of the particle satisfies $E>0$ and $L>0$, it is evident
that circular time-like geodesics exist only when $r>r_{ps}^{+}$,
$r<r_{ps}^{-}$ and $r>r_{*}=Q^{2}/M.$ In contrast, as photon spheres
do not exist $(Q/M>\sqrt{9/8})$, circular time-like geodesics exist
when $r>r_{*}.$ Besides, the existence of stable circular orbits
is also bounded by the Marginal Stable Circular Orbits (MSCOs) with
the additional condition 
\begin{equation}
V_{\text{eff}}^{^{\prime\prime}}\left(r\right)=0.\label{eq:MSCO2}
\end{equation}
This condition $\left(\ref{eq:MSCO2}\right)$ yields a single solution
$r_{\text{msco}}$, two solutions $r_{\text{msco}}^{\pm}$ and no
solution for a RN black hole with $Q/M<1$, a RN naked singularity
with $1<Q/M<\sqrt{5}/2$ and a RN naked singularity with $Q/M>\sqrt{5}/2$,
respectively. Combining these findings with the results derived from
eqns. $\left(\ref{eq:MSCO1}\right)$ regarding the existence of circular
orbits, we collate four parameter spaces with distinct regions of
stable circular orbits, as outlined in the Table. \ref{Orbit Class}.

\begin{table}
{\scriptsize{}}%
\begin{tabular}{c|c|c}
\hline 
{\scriptsize{}$Q/M$ Value} & {\scriptsize{}Region of stable circular orbit} & {\scriptsize{}Photon Sphere}\tabularnewline
\hline 
\hline 
{\scriptsize{}$Q/M<1$} & {\scriptsize{}$r>r_{\text{msco}}$} & \multirow{2}{*}{{\scriptsize{}Existent}}\tabularnewline
\cline{1-2} \cline{2-2} 
{\scriptsize{}$1<Q/M<\sqrt{9/8}$} & {\scriptsize{}$r_{*}<r<r_{ps}^{-}$ and $r>r_{\text{\text{msco}}}^{+}$} & \tabularnewline
\hline 
{\scriptsize{}$\sqrt{9/8}<Q/M<\sqrt{5}/2$} & {\scriptsize{}$r_{*}<r<r_{\text{\text{msco}}}^{-}$ and $r>r_{\text{\text{msco}}}^{+}$} & \multirow{2}{*}{{\scriptsize{}Non-Existent}}\tabularnewline
\cline{1-2} \cline{2-2} 
{\scriptsize{}$Q/M>\sqrt{5}/2$} & {\scriptsize{}$r>r_{*}$} & \tabularnewline
\hline 
\end{tabular}{\scriptsize\par}

\caption{Region of stable circular orbit and photon sphere. The analytical
expressions for $r_{\text{\text{msco}}}$ and $r_{\text{\text{msco}}}^{\pm}$
are given in \citep{Pugliese:2010ps}. }

\label{Orbit Class}
\end{table}

In the numerical simulation presented below, our primary focus lies
in comparing the simulation results for $Q/M=0.9,1.05$ and $1.11$.
These specific values are chosen to represent distinct scenarios,
namely, black holes, naked singularities with a photon sphere, and
naked singularities without a photon sphere, respectively.

\section{Celestial Sphere Model}

\label{sec:Celestial-Sphere-Model}

\begin{figure}
\includegraphics[scale=0.051]{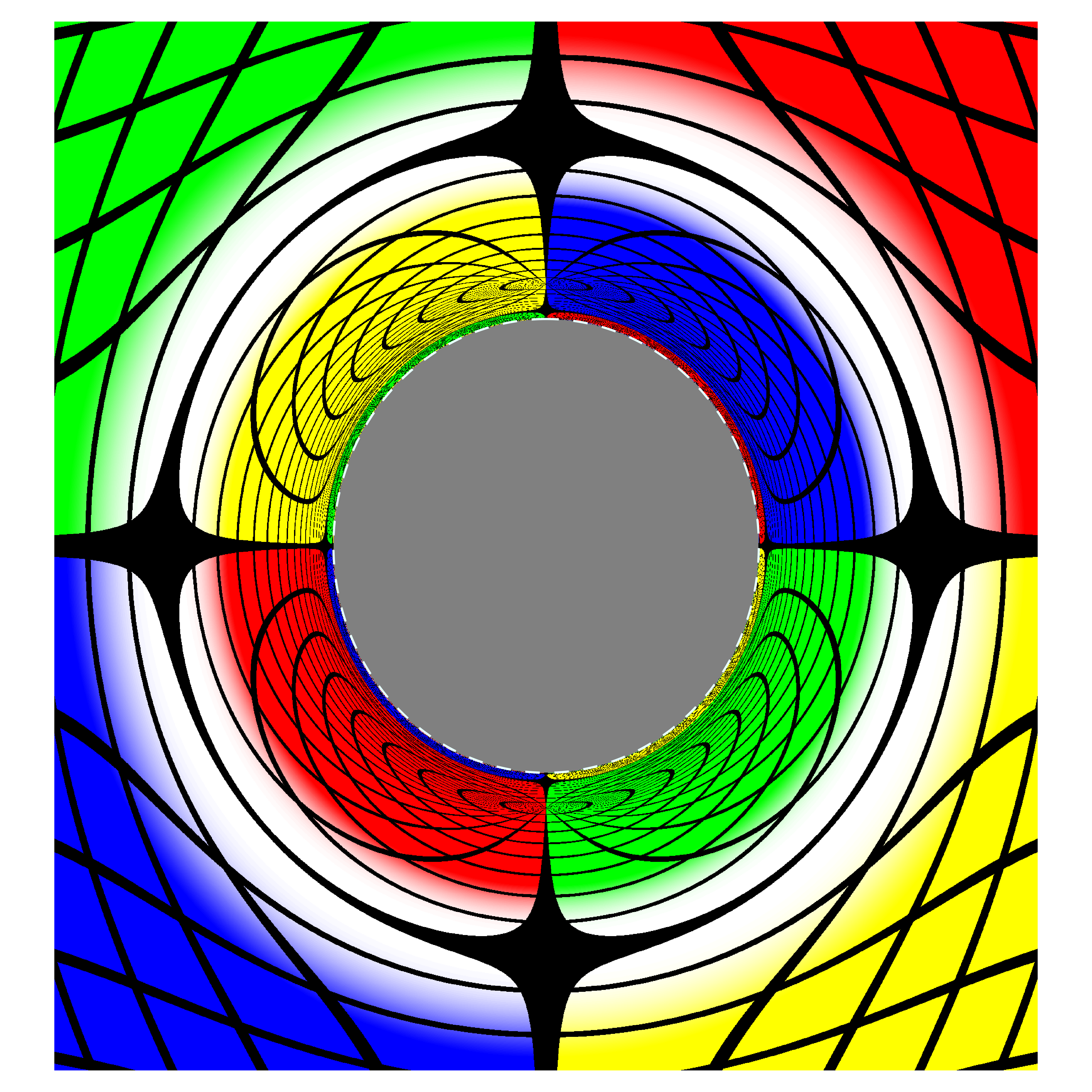} \includegraphics[scale=0.051]{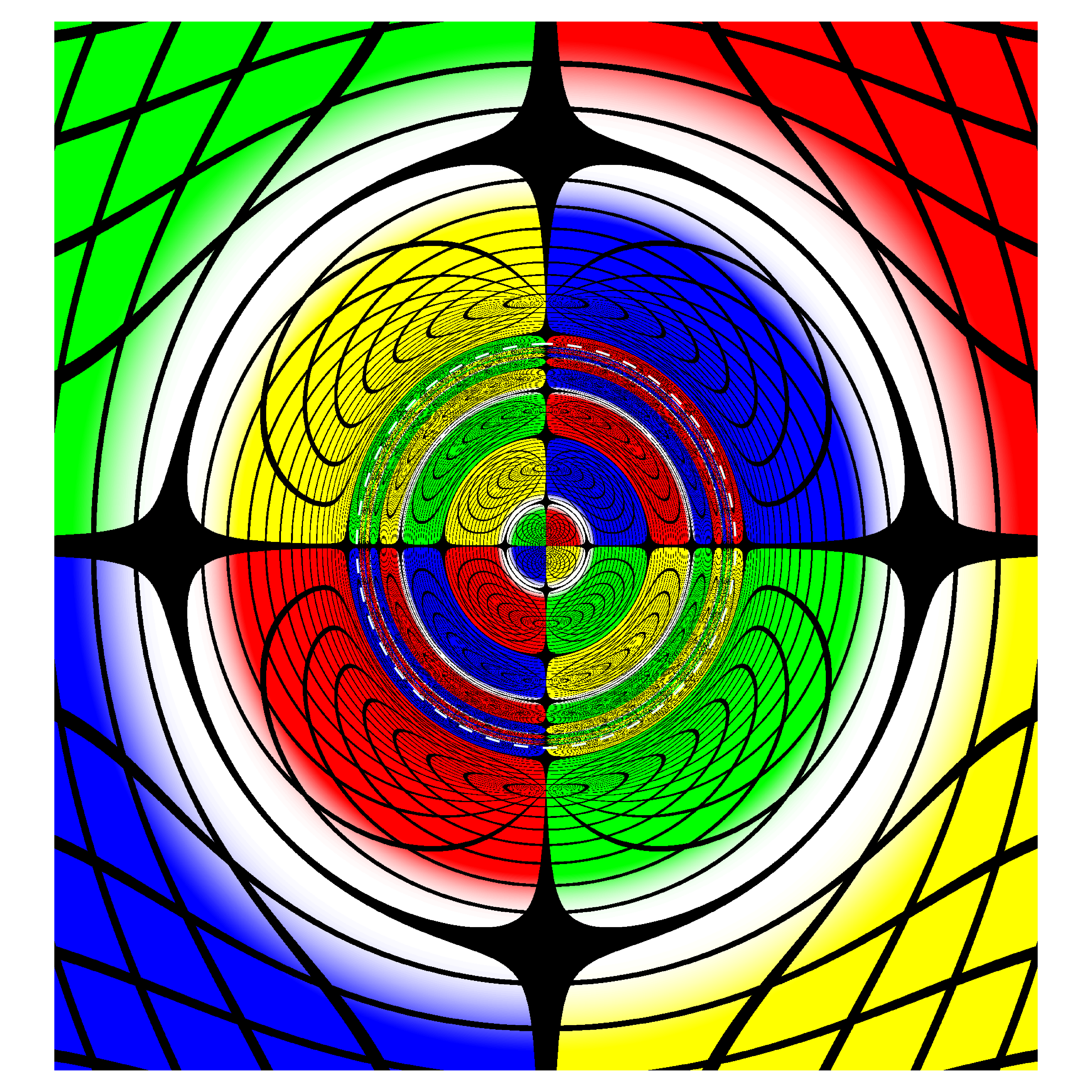}
\includegraphics[scale=0.051]{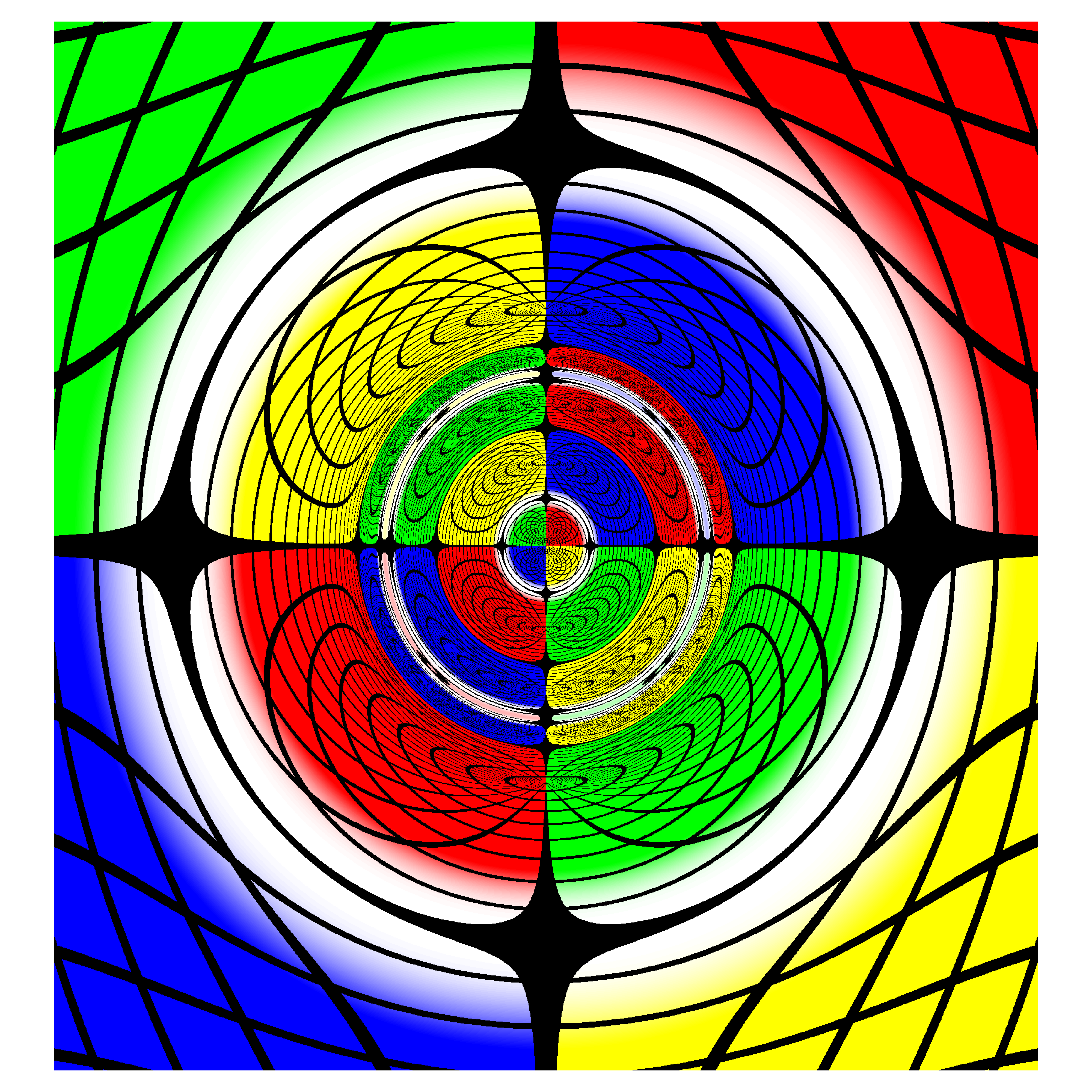}\caption{Images of a celestial sphere located at $r=25M$ in the RN metric
with $Q/M=0.9$ (\textbf{Left}), $1.05$ (\textbf{Middle}) and $1.1$
(\textbf{Right}). The observer's location is $x_{\text{o}}^{\mu}=(0,10M,\pi/2,\pi)$,
and the field of view spans $2\pi/5$. Dashed lines represent the
critical curve, formed by light rays emitted from the photon sphere.
The left panel exhibits one Einstein ring, while the middle and right
panels display three Einstein rings. In the case of $Q/M=0.9$, corresponding
to a black hole, the image encompasses the black hole shadow portrayed
as a gray area in the left panel. The middle panel illustrates the
image of a naked singularity with a photon sphere, where the region
outside the critical curve shows little deviation from the left panel.
However, within the critical curve, additional celestial images are
also present. Moreover, near the critical curve, there exist infinitely
high-order images. For the right panel, lacking a critical curve due
to the absence of a photon sphere, the overall pattern is similar
to the middle panel. Nevertheless, finite high-order images are observed
near the second Einstein ring.}
\label{fig:CS}
\end{figure}

This section utilizes numerical simulation to visualize gravitational
lensing around RN spacetime for an intuitive understanding of light
propagation. We model a celestial sphere source, divided into four
color-coded quadrants, with a white dot placed in front of the observer.
Additionally, a grid of black lines, spaced at $\pi/18$ intervals,
overlays the image to indicate constant longitude and latitude. To
generate observational images, we vary the observer\textquoteright s
viewing angle and numerically integrate $2000\times2000$ photon trajectories
until they intersect with the celestial sphere. For a detailed explanation
of the numerical implementation, interested readers can refer to \citep{Chen:2023trn}.

FIG. \ref{fig:CS} depicts images of the celestial sphere in both
RN black holes and singularities. Dashed circular lines depicted the
critical curves, shaped by light rays escaping the photon spheres.
The left panel displays the celestial sphere image for $Q/M=0.9$.
It includes the black hole shadow rendered in gray, an Einstein ring,
and some higher-order images visible outside the critical curve.

The middle panel of FIG. \ref{fig:CS} depicts the celestial sphere
image for a RN naked singularity of $Q/M=1.05$, possessing a photon
sphere. Similar to the $Q/M=0.9$ case, the region outside the critical
curve shows a comparable appearance. However, light ray refraction
within the photon sphere leads to additional celestial images appearing
inside the critical curve. Furthermore, compression and color alternation
near the critical curve indicate the generation of numerous higher-order
terms in its vicinity.

The right panel of FIG. \ref{fig:CS} displays the celestial sphere
image for a RN naked singularity of $Q/M=1.1$, devoid of photon spheres.
This image resembles the middle panel at a distance from the critical
curve. However, the infinitely higher-order images converging near
the critical curve are replaced by a finite number of higher-order
images.

\begin{figure}
\includegraphics[scale=0.9]{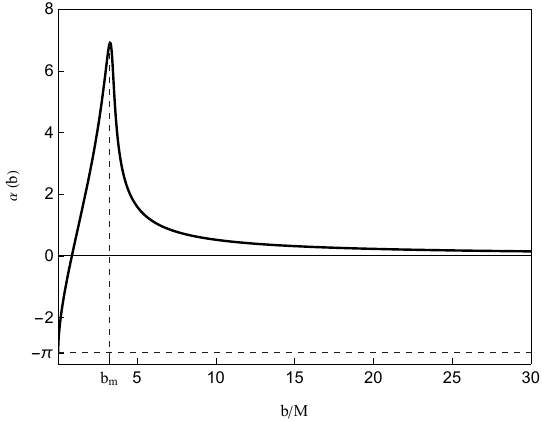} \caption{The deflection angle $\alpha(b)$ as a function of the impact parameter
$b$ for $Q/M=1.1$. The deflection angle reaches its maximum at $b=b_{m}$.}
\label{fig:deflection angle}
\end{figure}

To understanding the finite number of higher-order images in the case
of $Q/M=1.1$, we analyze the deflection angle as a function of the
impact parameter $b$, depicted FIG. \ref{fig:deflection angle}.
In this figure, the deflection angle is defined as $\alpha(b)=I(b)-\pi$,
where the $I(b)$ is numerically integrated (details in \citep{Chen:2023trn,Chen:2023uuy}).
Notably, as the impact parameter $b$ varies from $0$ to the infinity,
the deflection angle exhibits a maximum at $b_{m}$, leading to the
absence of infinitely higher-order images.

\section{Hot Spot Model}

\label{sec:Hot Spot Model}

This section addresses the observational characteristics exhibited
by hot spots surrounding RN black holes and naked singularities. Consistent
with available data and theoretical analyses, flares likely originate
from synchrotron radiation emitted by matter near the innermost stable
circular orbit (ISCO) \citep{Hamaus:2008yw,Trippe:2006jy,Broderick:2005jj}.
To accurately predict flare properties, we maintain hot spot orbits
close to the ISCO. For the RN black hole with $Q/M=0.9$, a single
stable orbit region exists, so the MSCO with radius $r_{\text{msco}}$
is chosen. Conversely, RN naked singularities with $Q/M=1.05$ and
$1.1$ exhibit two distinct stable orbit regions (discussed in Sec.
\ref{sec:Set up}). We investigate hot spots in both outer and inner
regions. In the outer region, the outer MSCO with radius $r_{\text{\text{msco}}}^{+}$
is adopted as the orbit of the hot spot. In the inner region, an orbit
near radius $r^{*}$ is chosen, as $r=r^{*}$ implies a geodesic with
zero angular momentum. Specifically, for $Q/M=1.05$ and $1.1$, inner
orbits at $r=1.11M$ and $1.22M$, respectively, are designated for
the hot spots.

To simplify calculations, we model the hot spot as an isotropically
emitting sphere and utilize our computational framework described
in \citep{Hamaus:2008yw,Rosa:2022toh,Rosa:2023qcv,Chen:2023knf}.
Within this framework, the observer is positioned at $\left(t_{o},r_{o},\theta_{o},\varphi_{o}\right)=\left(t_{o},100M,\theta_{o},\pi\right)$.
For computational efficiency and precision, we use a $1000\times1000$
pixel grid for each snapshot and generate $500$ snapshots for a period
of the orbit, ensuring smooth image evolution throughout period $T_{e}$.
At a specific time $t_{k}$, each pixel in the image plane receives
an intensity $I_{klm}$, collectively forming lensed hot spot images.
We then analyze the following image properties, following \citep{Rosa:2022toh,Rosa:2023qcv},
\begin{itemize}
\item Time integrated image: 
\begin{equation}
\left\langle I\right\rangle _{lm}=\sum\limits _{k}I_{klm}.
\end{equation}
\item Total temporal flux: 
\begin{equation}
F_{k}=\sum\limits _{l}\sum\limits _{m}\Delta\Omega I_{klm},
\end{equation}
where $\Delta\Omega$ corresponds to the solid angle of a pixel.
\item Temporal magnitude: 
\begin{equation}
m_{k}=-2.5\lg\left(\frac{F_{k}}{\min\left(F_{k}\right)}\right).
\end{equation}
\item Temporal centroid: 
\begin{equation}
\overrightarrow{c_{k}}=F_{k}^{-1}\sum\limits _{l}\sum\limits _{m}\Delta\Omega I_{klm}\overrightarrow{r_{lm}},
\end{equation}
where $\overrightarrow{r_{lm}}$ represents the position relative
to the image center. 
\end{itemize}

\subsection{Integrated Images }

\label{subsec:Integrated-Images}

\begin{figure}[H]
\includegraphics[width=0.33\textwidth]{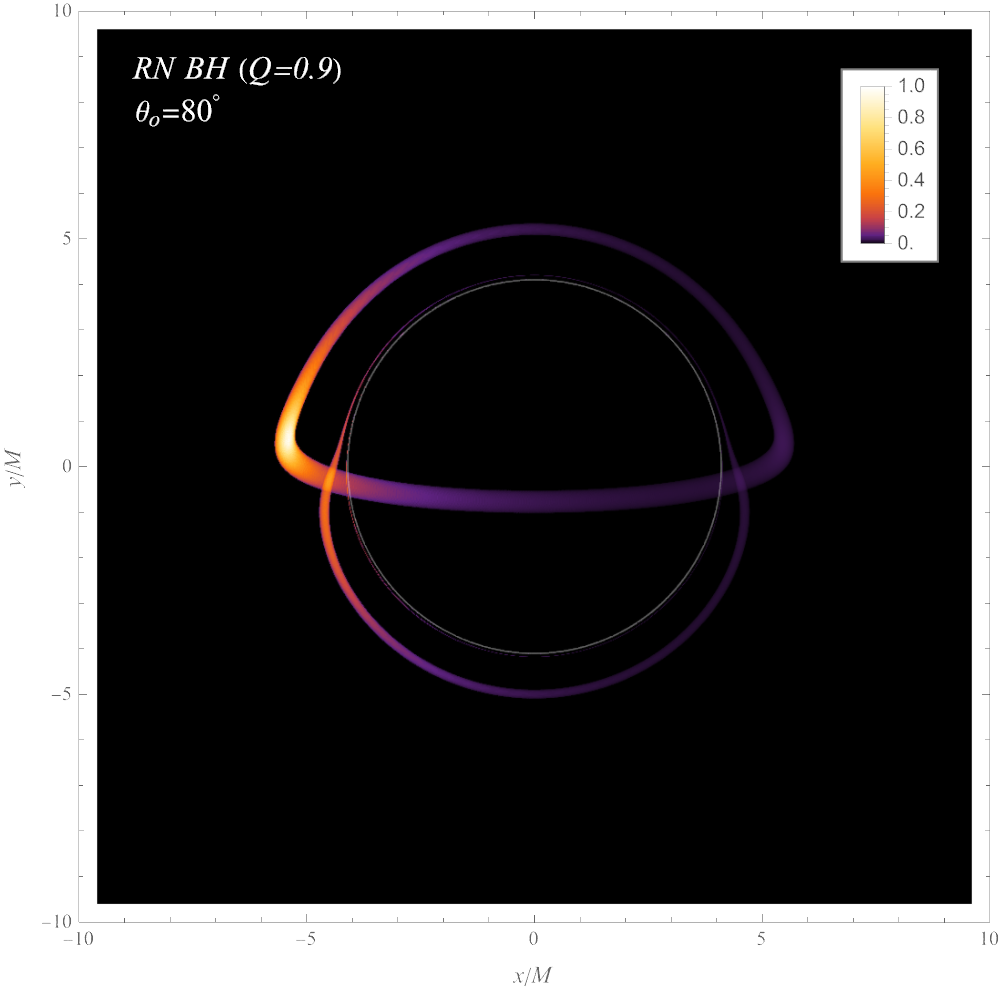}\includegraphics[width=0.33\textwidth]{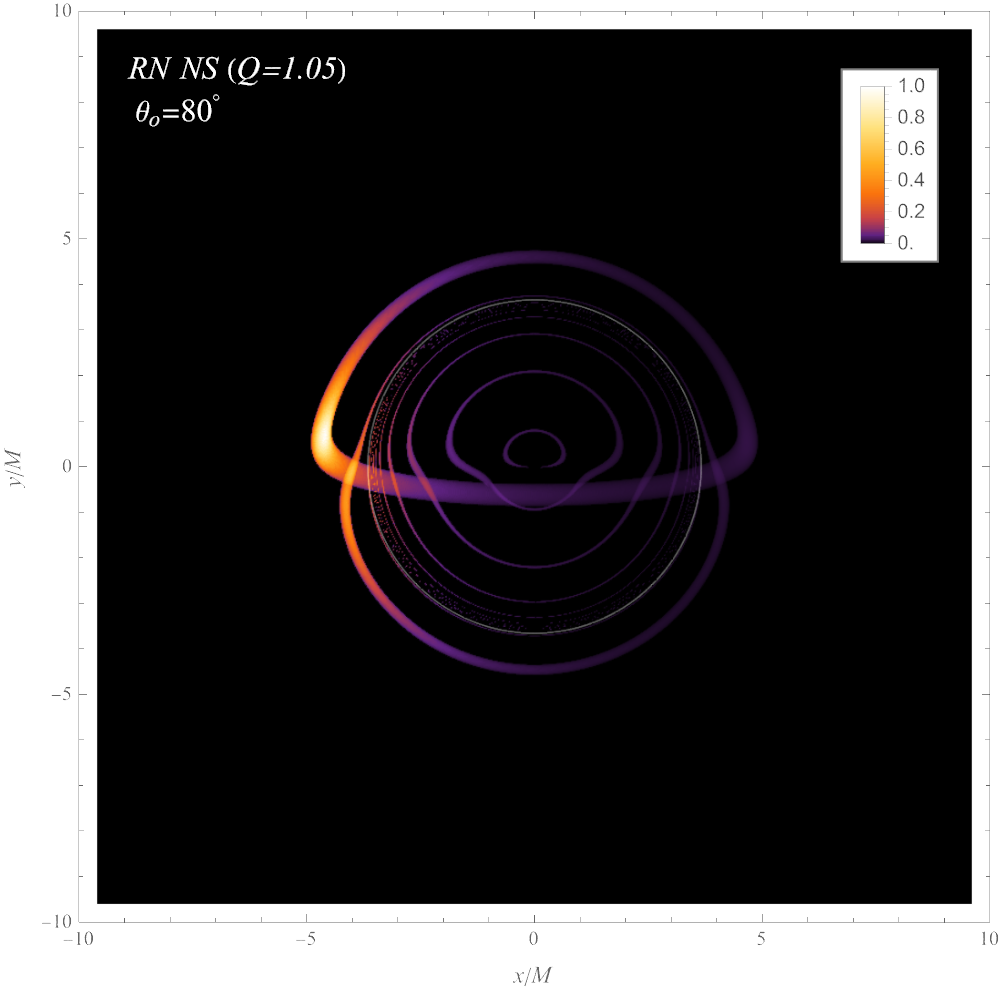}\includegraphics[width=0.33\textwidth]{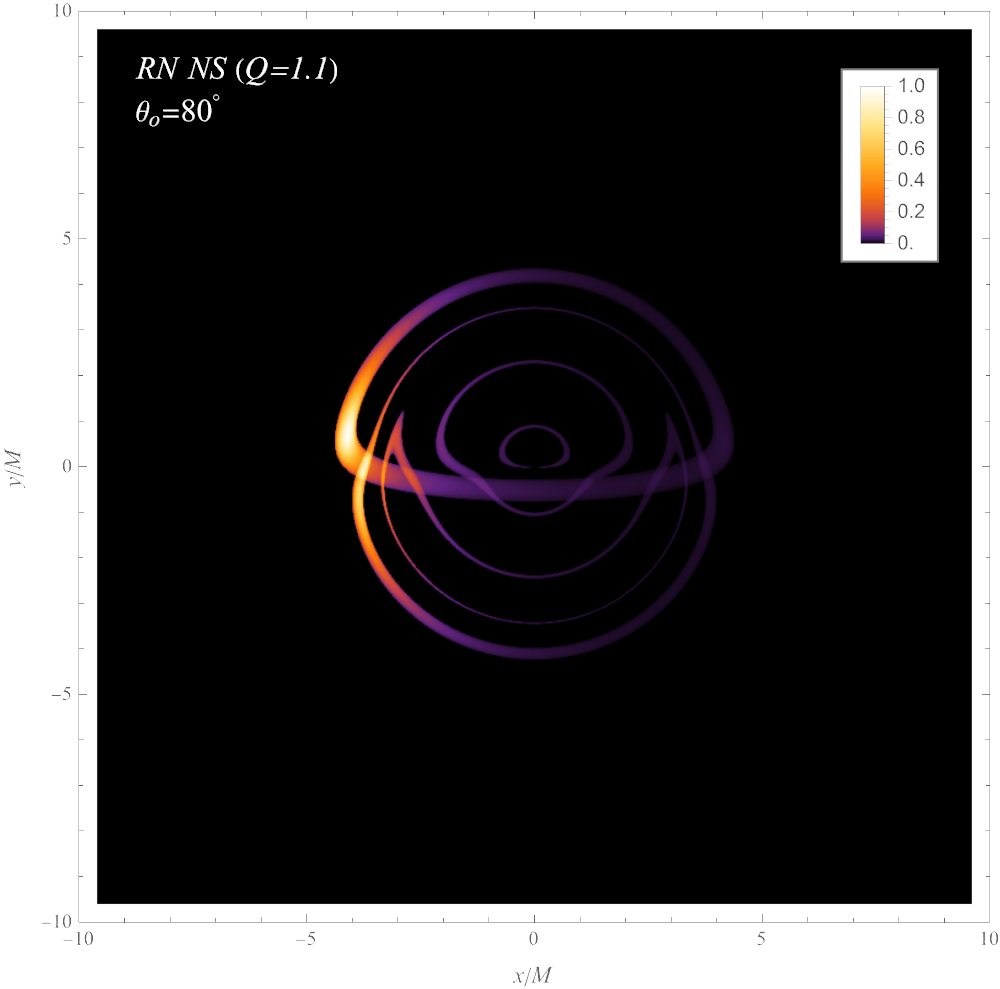}

\includegraphics[width=0.33\textwidth]{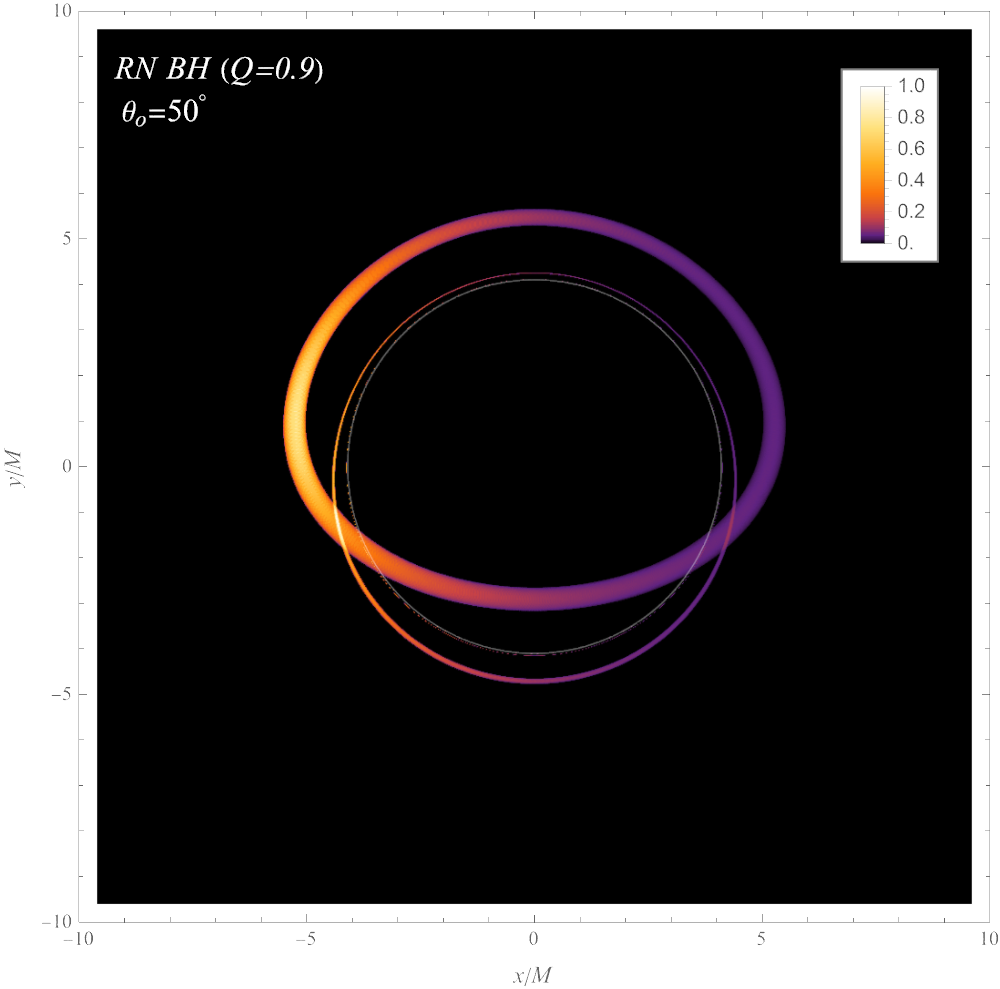}\includegraphics[width=0.33\textwidth]{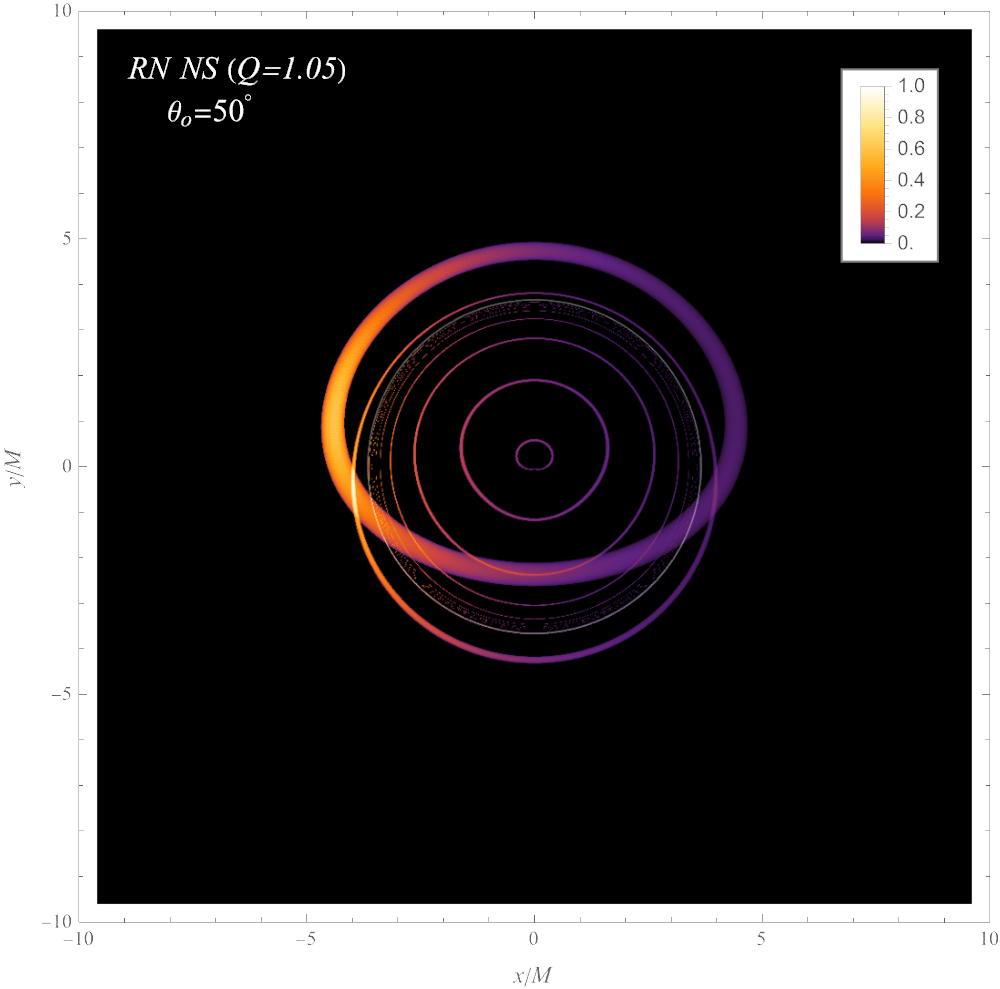}\includegraphics[width=0.33\textwidth]{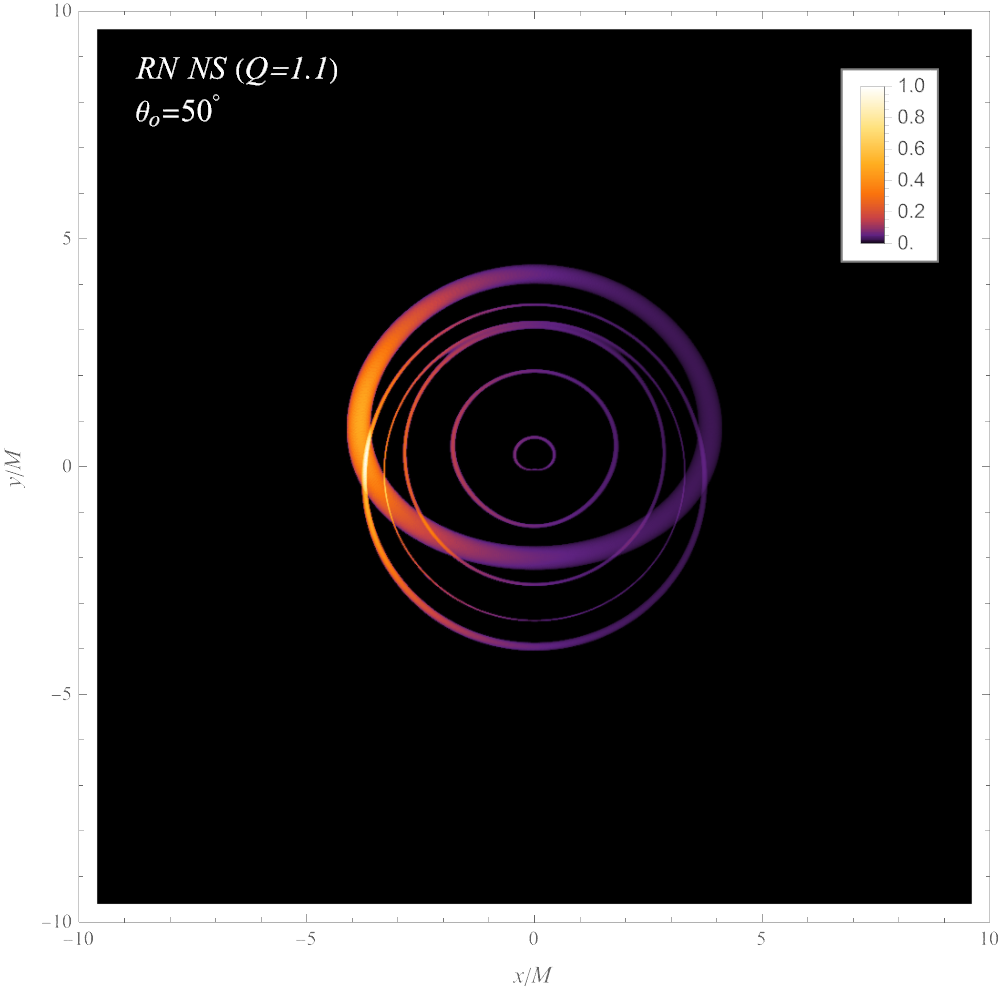}\caption{Time integrated images depicting a full orbit of the hot spot are
obtained from an observational inclination angle of $\theta_{o}=80^{\circ}$
(\textbf{Upper Row}) and $\theta_{o}=50^{\circ}$ (\textbf{Lower Row}).
Note that the intensity are normalized with respect to their maximum
value. The white lines delineate the critical curves, shaped by light
rays escaping from the photon spheres.\textbf{ Left Column}: RN black
hole with $Q/M=0.9$, and the hot spot is situated at $r=r_{\text{\text{msco}}}$.
This image showcases the primary and secondary lensed image tracks
positioned outside the critical curve, arising from the $n=0^{>}$
and $1^{>}$ light rays emitted by the hot spot, respectively. \textbf{Middle
Column}: RN singularity with $Q/M=1.05$, and the hot spot is situated
at $r=r_{\text{\text{msco}}}^{+}$. The image reveals two image tracks
located outside the critical curve, akin to the black hole case. Owing
to the reflection of light rays entering the photon sphere, additional
tracks emerge within the critical curve. \textbf{Right Column}: RN
singularity with $Q/M=1.1$, and the hot spot is situated at $r=r_{\text{\text{msco}}}^{+}$.
This image displays finite image tracks due to the absence of a photon
sphere. Moreover, the two tracks of $2^{<}$ and $2^{>}$merge together.}
\label{fig:Inte1}
\end{figure}

FIG. \ref{fig:Inte1} reveals the time integrated images for hot spots
orbiting around a RN black hole at $r=r_{\text{msco}}$ and RN naked
singularities at $r=r_{\text{msco}}^{+}$. The upper and lower panels
show images observed at inclination angles of $\theta_{o}=80^{\circ}$
and $50^{\circ}$, respectively. These images manifest several image
tracks revealing the orbit of hot spots. To comprehend the origin
of these tracks, we use the integer $n$ to indicate the number of
times light rays intersect the equatorial plane. Furthermore, in the
presence of a photon sphere in cases of the RN black hole with $Q/M=0.9$
and the RN naked singularity with $Q/M=1.05$, image tracks are separated
by the critical curve, shaped by light rays with the critical impact
parameter $b_{c}$. In the absence of a photon sphere ($Q/M=1.1$),
the deflection angle reaches a maximum value $b_{m}$ as the impact
parameter $b$ varies. We therefore introduce superscripts $>$ and
$<$ to distinguish light rays with $b>b_{c}\left(b_{m}\right)$ and
$b<b_{c}\left(b_{m}\right)$, respectively.

The left column of FIG. \ref{fig:Inte1} displays time integrated
images of a hot spot orbiting an RN black hole with $Q/M=0.9$ at
$r=r_{\text{\text{msco}}}$. These images reveal a central shadow
surrounded by a critical curve. For an observer's inclination angle
of $\theta_{o}=80^{\circ}$, the primary image $\left(n=0^{>}\right)$
exhibits a closed, semicircular track. Its upper and lower segments
correspond to the hot spot behind and in front of the black hole,
respectively. In contrast, the secondary image with $n=1^{>}$ follows
a smaller and fainter track. Moreover, images of higher order have
significantly reduced luminosity and adhere closely to the critical
curve. Additionally, these images exhibit a distinctive brightness
asymmetry due to Doppler effects.

The image obtained at an observation inclination of $\theta_{o}=50^{\circ}$
resembles the one at $\theta_{o}=80^{\circ}$, but with a lower level
of brightness asymmetry observed at the lower inclination. Furthermore,
the image tracks are more circular at the lower inclination.

The middle column of FIG. \ref{fig:Inte1} shows the time integrated
image of a hot spot orbiting the RN singularity with $Q/M=1.05$ at
$r=r_{\text{\text{msco}}}^{+}$. There are two tracks outside the
critical curve, resembling the scenario for the black hole at both
observation inclinations of $\theta_{o}=50^{\circ}$ and $80^{\circ}$.
In contrast, four distinct tracks ($n=0^{<}$, $1^{<},$ $2^{<}$
and $3^{<}$) appear within the critical curve, generated by light
rays refracted around the naked singularity. Similarly, as the order
$n$ increases, the track becomes progressively more circular, exhibits
significantly reduced luminosity, and closely follows the critical
curve.

Unlike the previous cases, the absence of a photon sphere in the RN
singularity with $Q/M=1.1$ results in no critical curve within the
time-integrated images for a hot spot orbiting at $r=r_{\text{\text{msco}}}^{+}$
(FIG. \ref{fig:Inte1}, right column). Instead, six distinct tracks
are visible, denoted as $n=0^{<}$, $1^{<}$, $2^{<}$, $2^{>}$,
$1^{>}$ and $0^{>}$. Notably, the $2^{<}$ and $2^{>}$ tracks appear
incomplete and converge when observed at $\theta_{o}=80^{\circ}$,
where their deflection angles $\alpha(b)$ reach their maximum. This
merging behavior persists at an inclination angle of $\theta_{o}=50^{\circ}$,
although both $2^{<}$ and $2^{>}$ tracks seem complete in this case.

\begin{figure}
\includegraphics[width=0.33\textwidth]{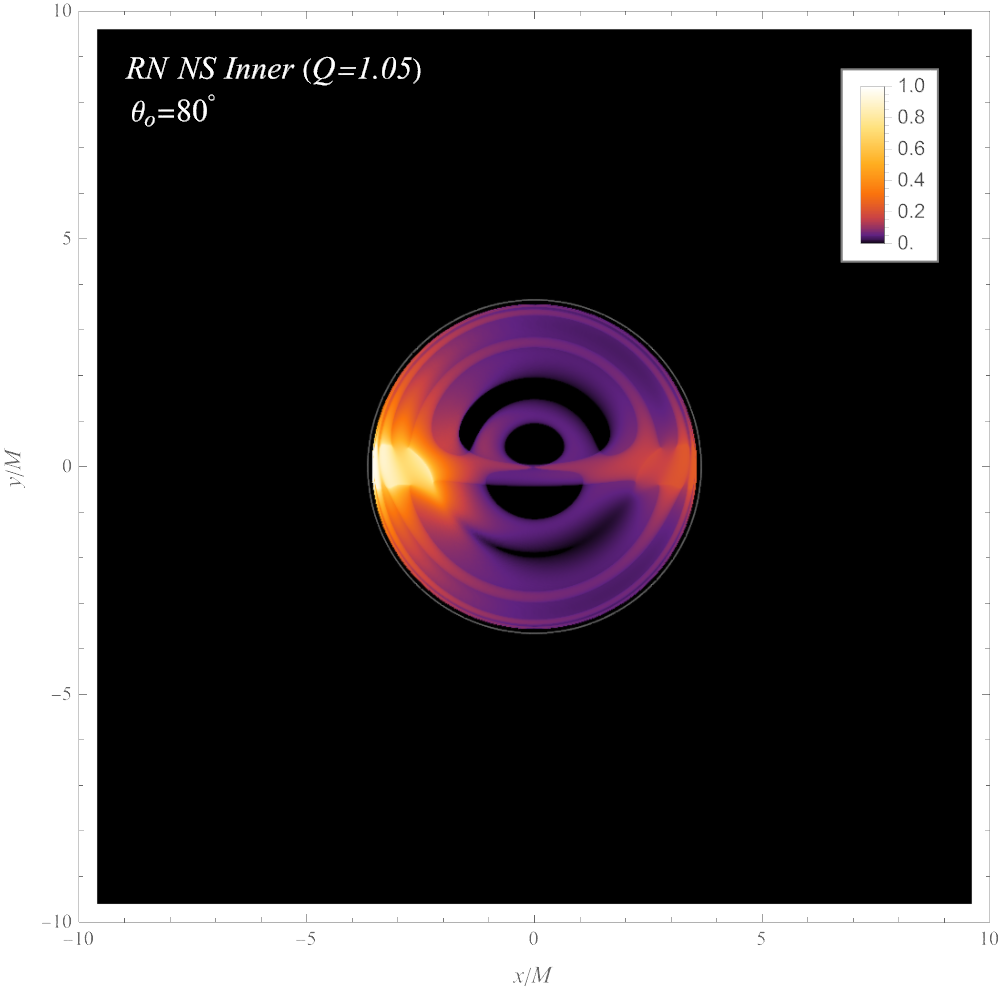}\includegraphics[width=0.33\textwidth]{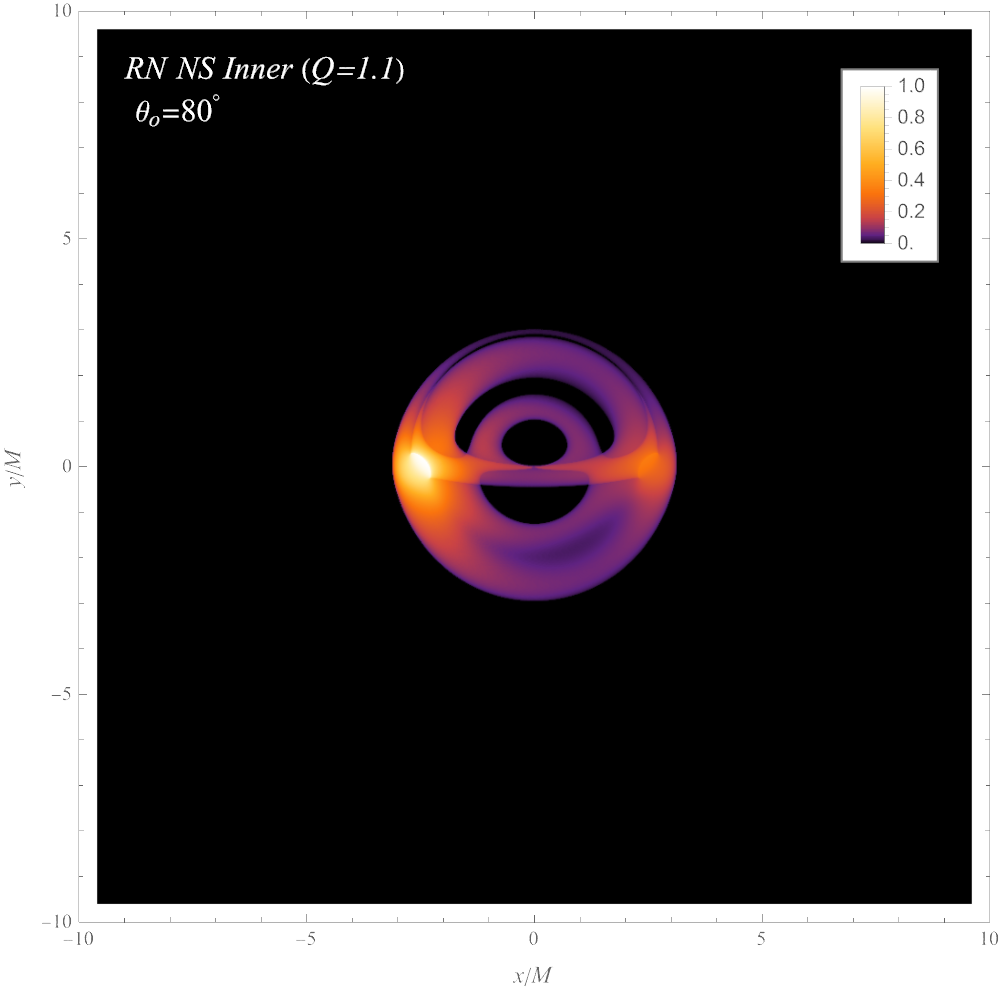}

\includegraphics[width=0.33\textwidth]{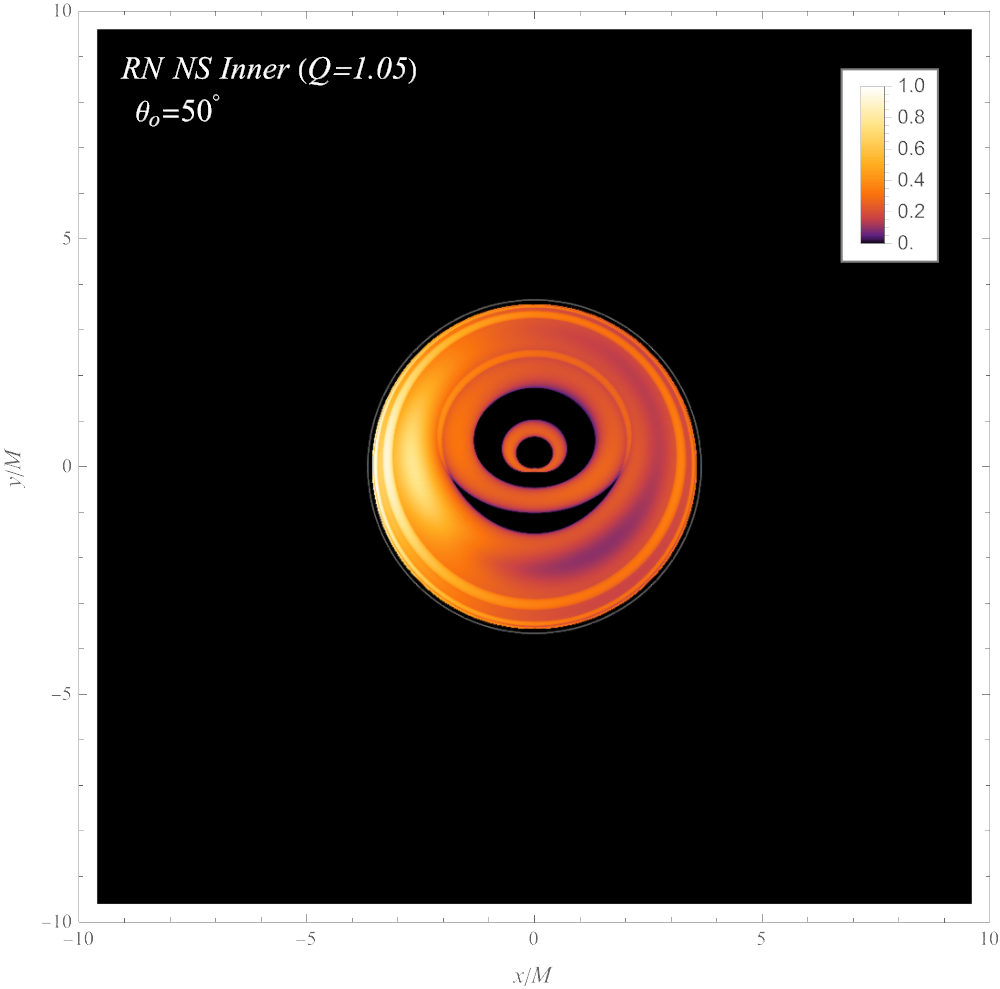}\includegraphics[width=0.33\textwidth]{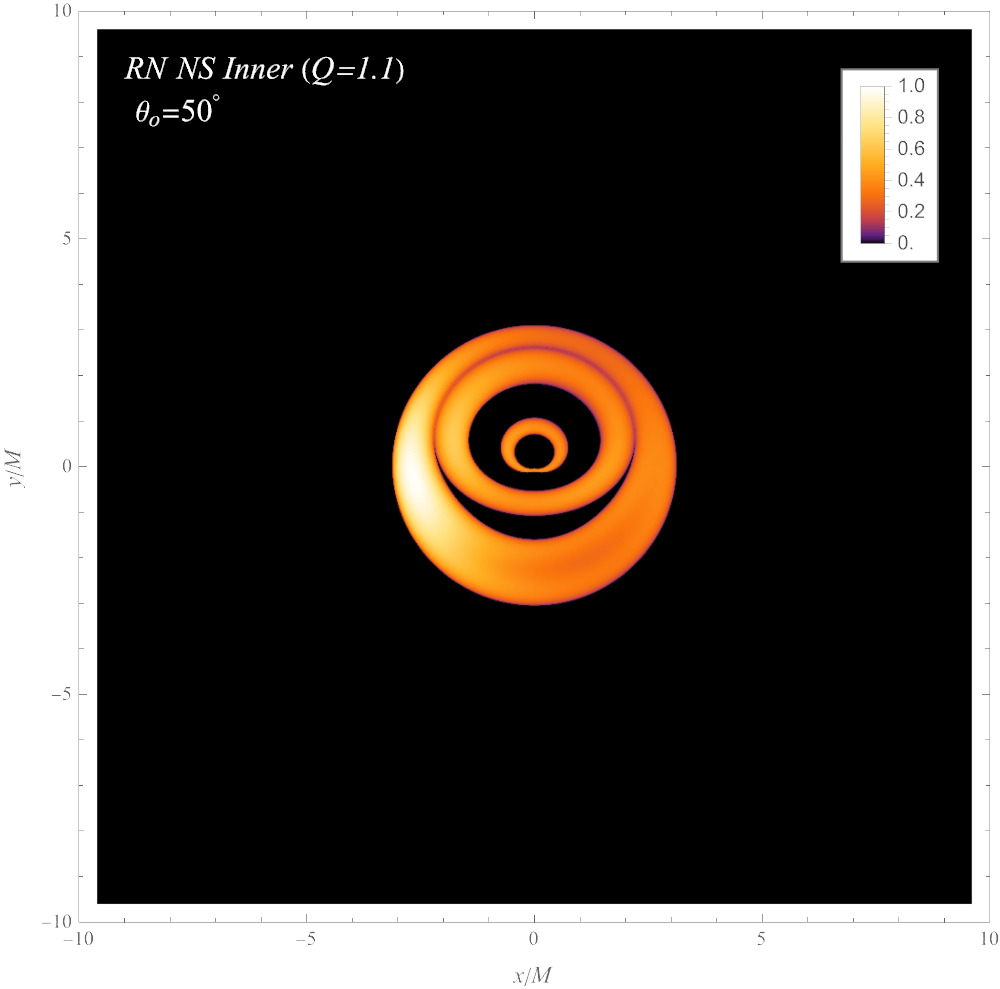}\caption{Time integrated images depicting a full orbit of the hot spot positioned
near $r=r^{*}$ in the RN singularity with $Q/M=1.05$ (\textbf{Left
Column}) and $1.1$ (\textbf{Right Column}). These images are obtained
from an observational inclination angle of $\theta_{o}=80^{\circ}$
(\textbf{Upper Row}) and $\theta_{o}=50^{\circ}$ (\textbf{Lower Row}).\textbf{
}As a result of the reduced orbit radius, the resulting images are
displayed in a smaller field of view. The tracks in the images appear
overlapped and are challenging to distinguish.}
\label{Integrated2}
\end{figure}

FIG. \ref{Integrated2} depicts integrated images of RN singularities
with $Q/M=1.05$ and $Q/M=1.1$, where the hot spots are orbiting
at $r=1.11M$ and $1.22M$, respectively. Due to their close proximity
to the center, the light rays experience a significant gravitational
influence, leading to the overlapping of final image tracks, making
them challenging to differentiate. For $Q/M=1.05$, the orbit of the
hot spot lies within the photon sphere, resulting in images bounded
by the critical curve. Besides, higher-order images are dimly discerned
near the critical curve, overlapped with the primary image. Conversely,
for $Q/M=1.1$, finite image tracks are observable, due to the absence
of a photon sphere. 

\subsection{Temporal Fluxes and Centroids}

\begin{figure}
\includegraphics[width=0.25\textwidth]{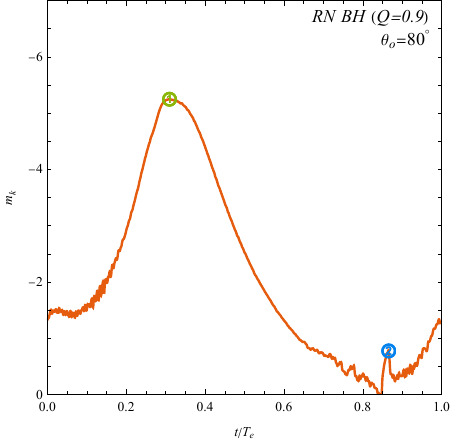}\includegraphics[width=0.25\textwidth]{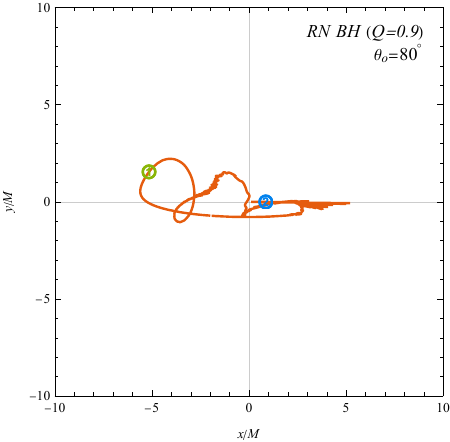}\includegraphics[width=0.25\textwidth]{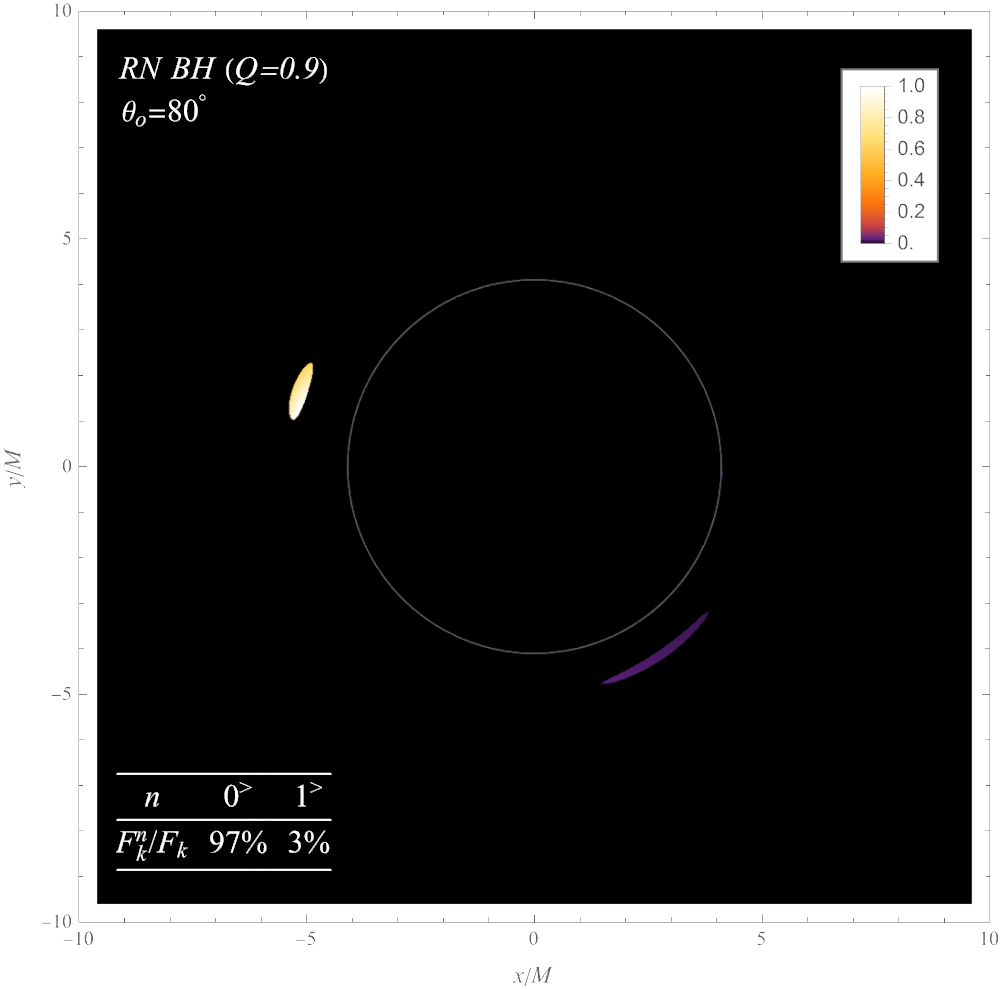}\includegraphics[width=0.25\textwidth]{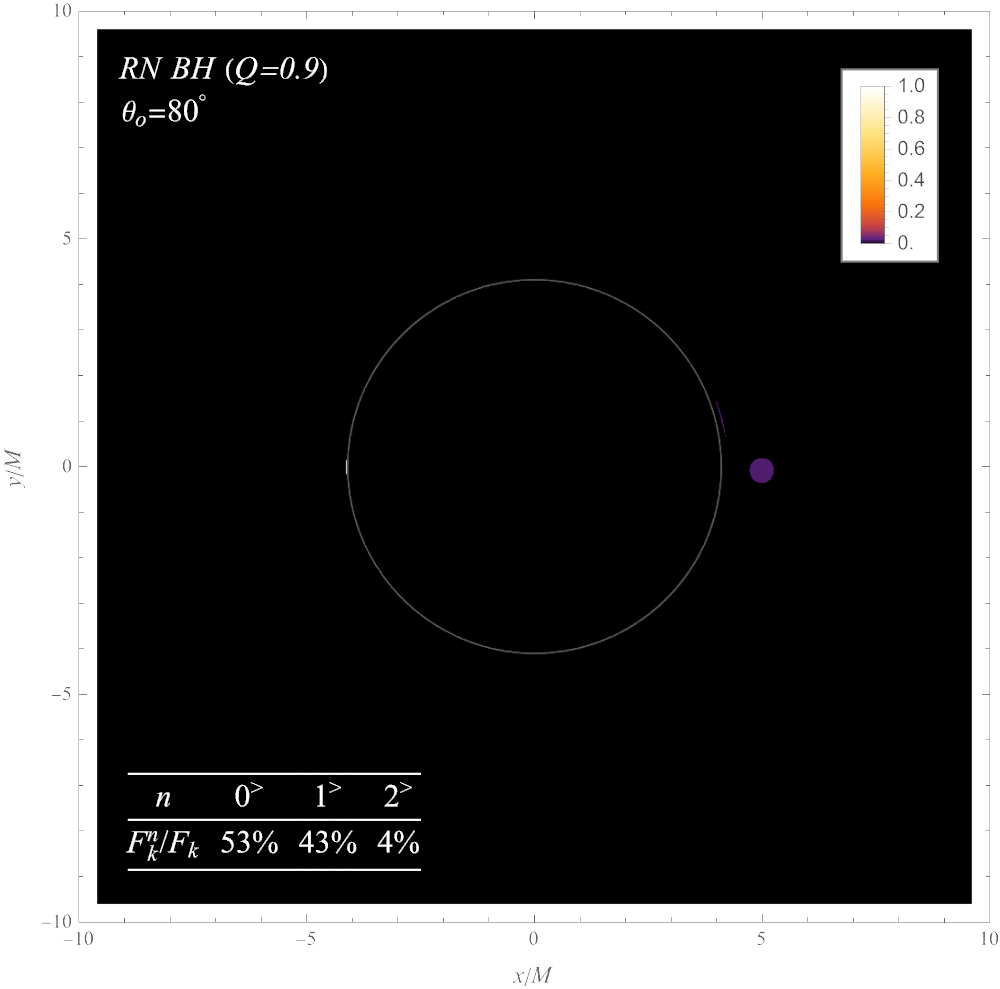}

\includegraphics[width=0.25\textwidth]{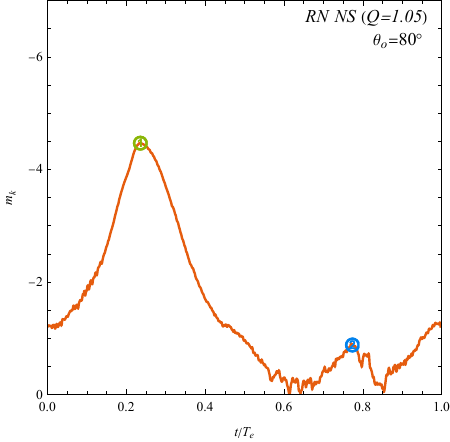}\includegraphics[width=0.25\textwidth]{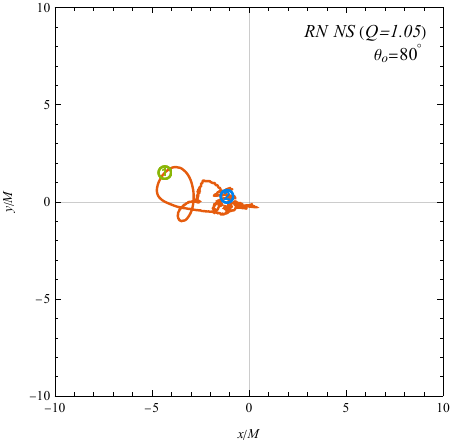}\includegraphics[width=0.25\textwidth]{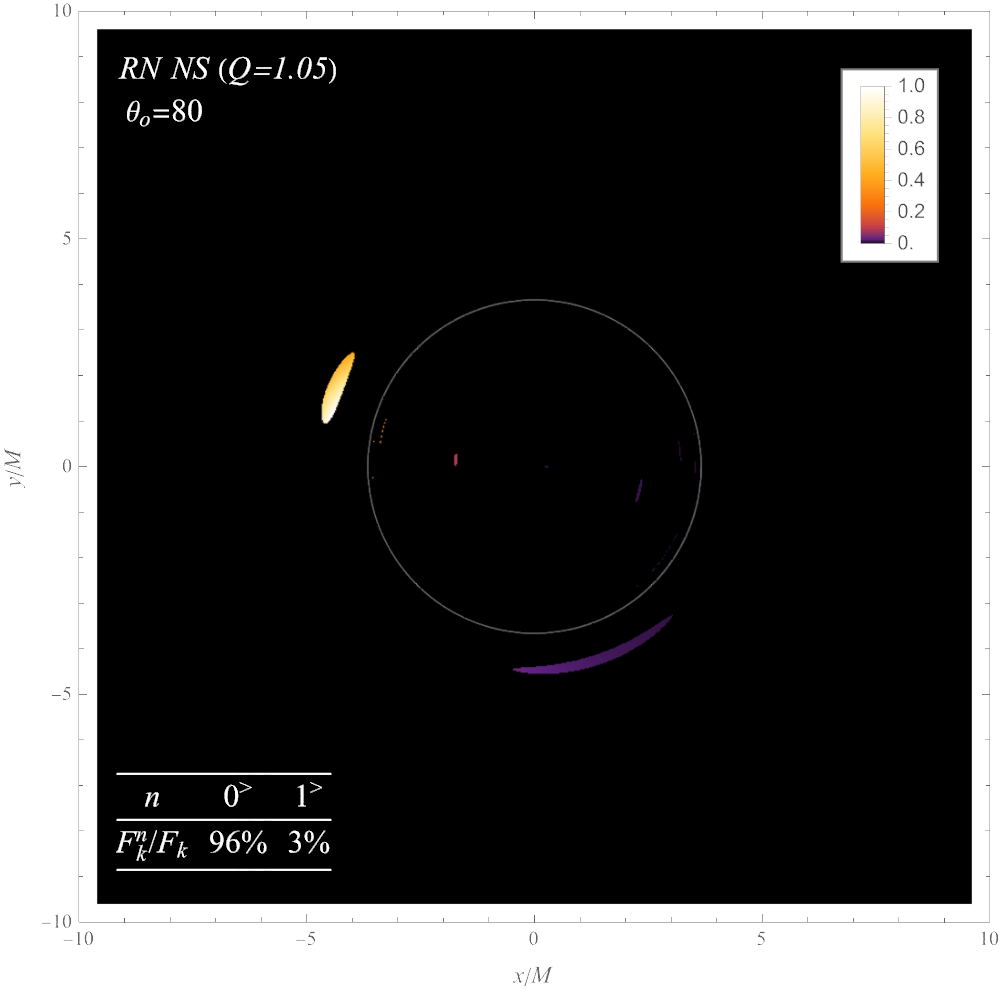}\includegraphics[width=0.25\textwidth]{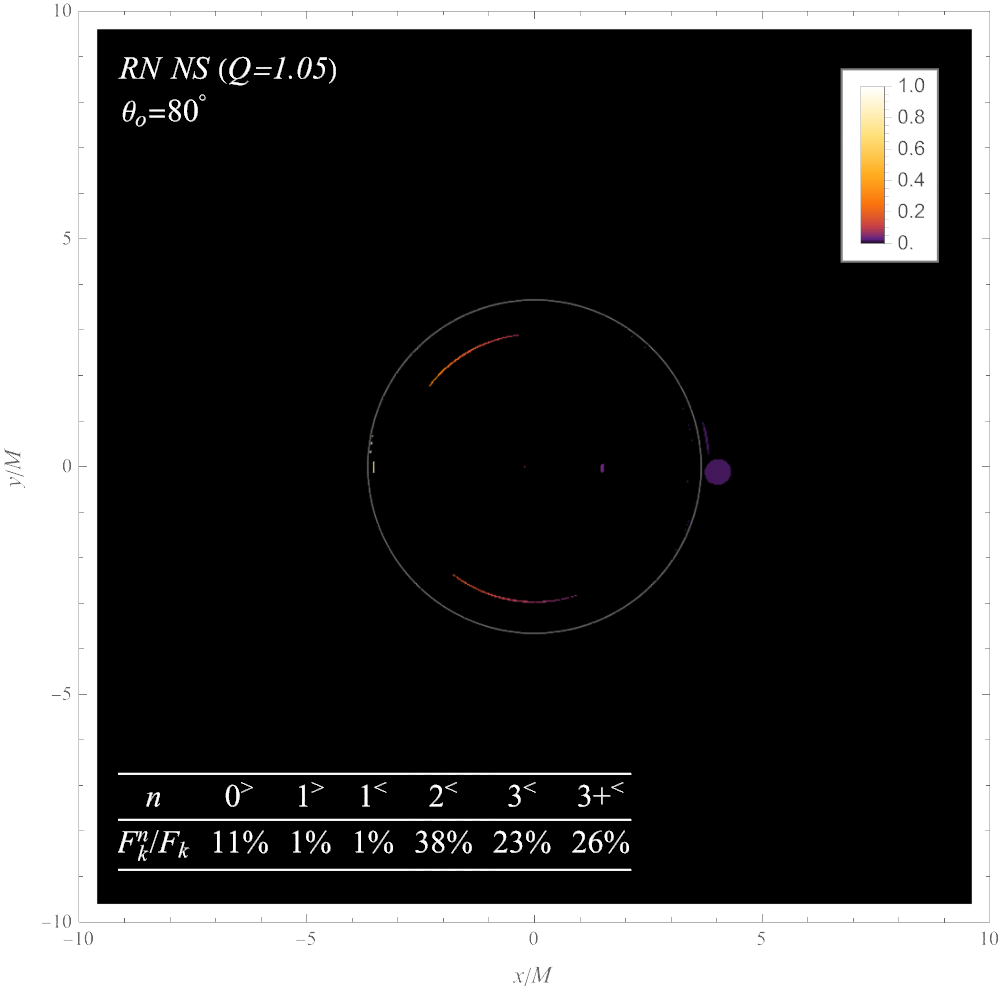}

\includegraphics[width=0.25\textwidth]{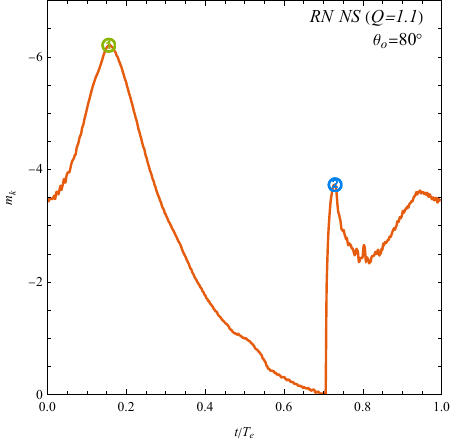}\includegraphics[width=0.25\textwidth]{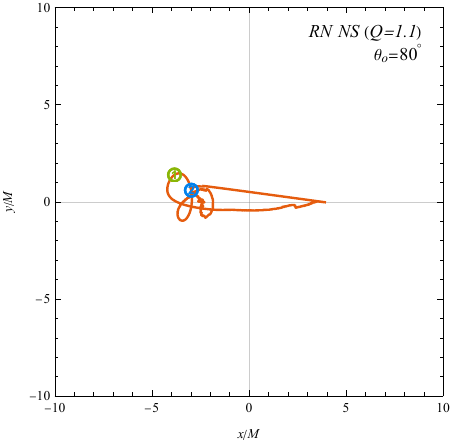}\includegraphics[width=0.25\textwidth]{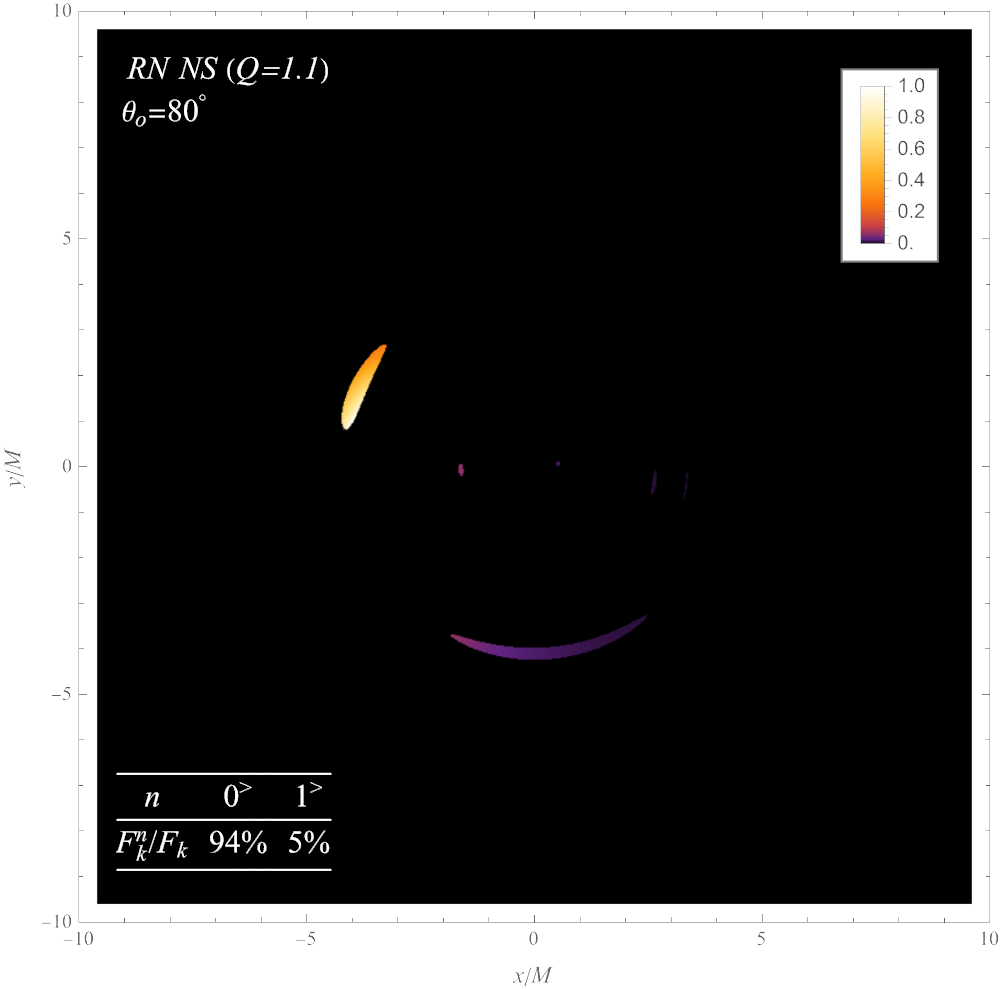}\includegraphics[width=0.25\textwidth]{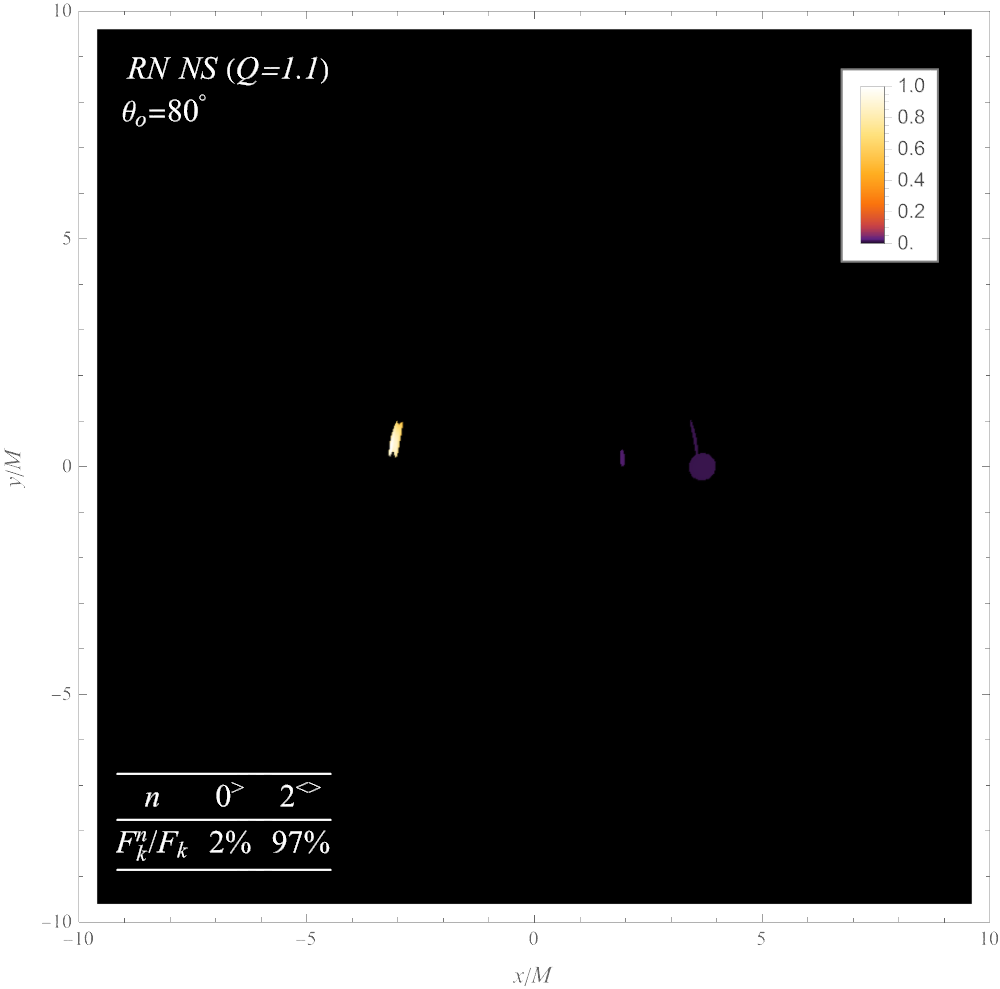}

\caption{Temporal magnitudes $m_{k}$ (\textbf{First Column}) and centroids
$c_{k}$ (\textbf{Second Column}) as a function of $t/T_{e}$ for
cases $Q/M=0.9$ with the orbit at $r_{\text{msco}}$ (\textbf{Top
Row}), $Q/M=1.05$ (\textbf{Middle Row}) and $Q/M=1.1$ with the orbit
at $r_{\text{msco}}^{+}$ (\textbf{Bottom Row}). The inclination is
$\theta_{o}=80^{\circ}$. The highest and second-highest peaks of
the temporal magnitude are indicated by \textcolor{darkpastelgreen}{\textcircled{1}}
and \textcolor{denim}{\textcircled{2}}, respectively. The centroids
of the flux at these peaks are identified with corresponding numbers.
The snapshots for each cases, when the temporal magnitude reaches
its highest peak and second-highest peak are presented in the third
and fourth column, respectively. The contribution from the $n^{\text{th}}$-order
image to the total flux is quantified by $F_{k}^{n}/F_{k}$, where
$F_{k}^{n}$ represents the temporal flux of the $n^{\text{th}}$-order
image at $t=t_{k}$.}
\label{fig:snapshots new1}
\end{figure}

\begin{figure}
\includegraphics[width=0.25\textwidth]{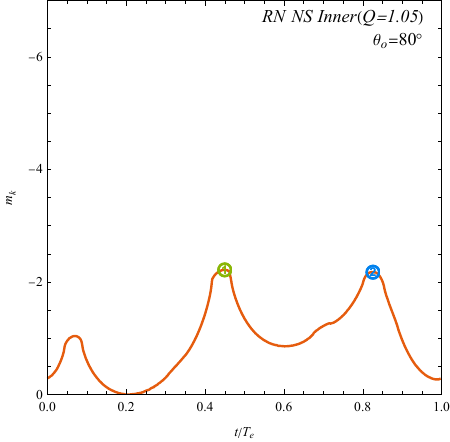}\includegraphics[width=0.25\textwidth]{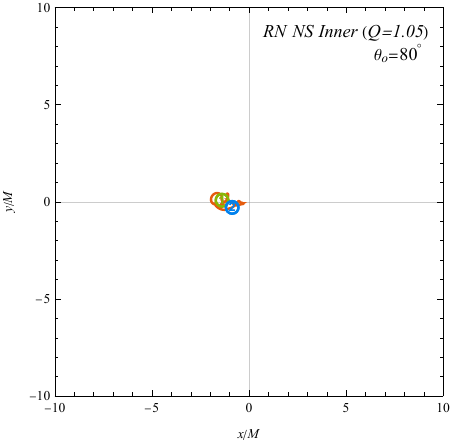}\includegraphics[width=0.25\textwidth]{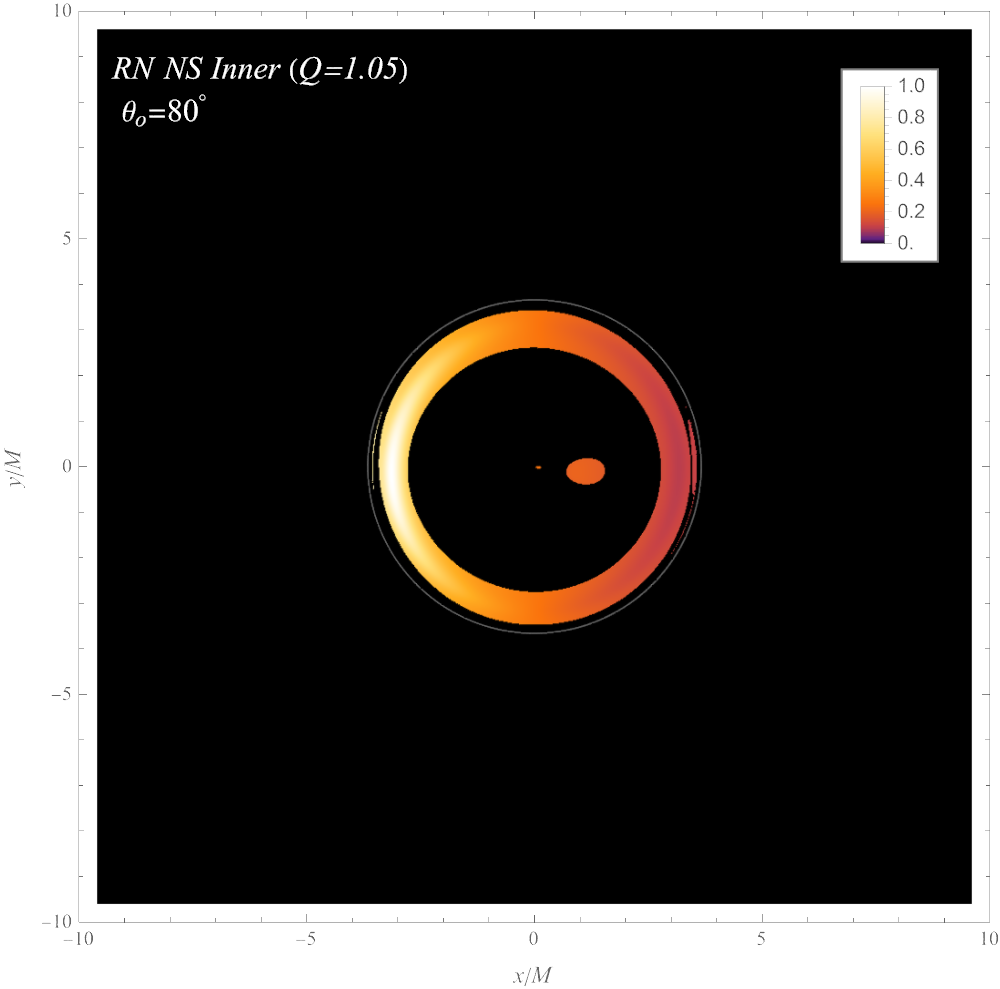}\includegraphics[width=0.25\textwidth]{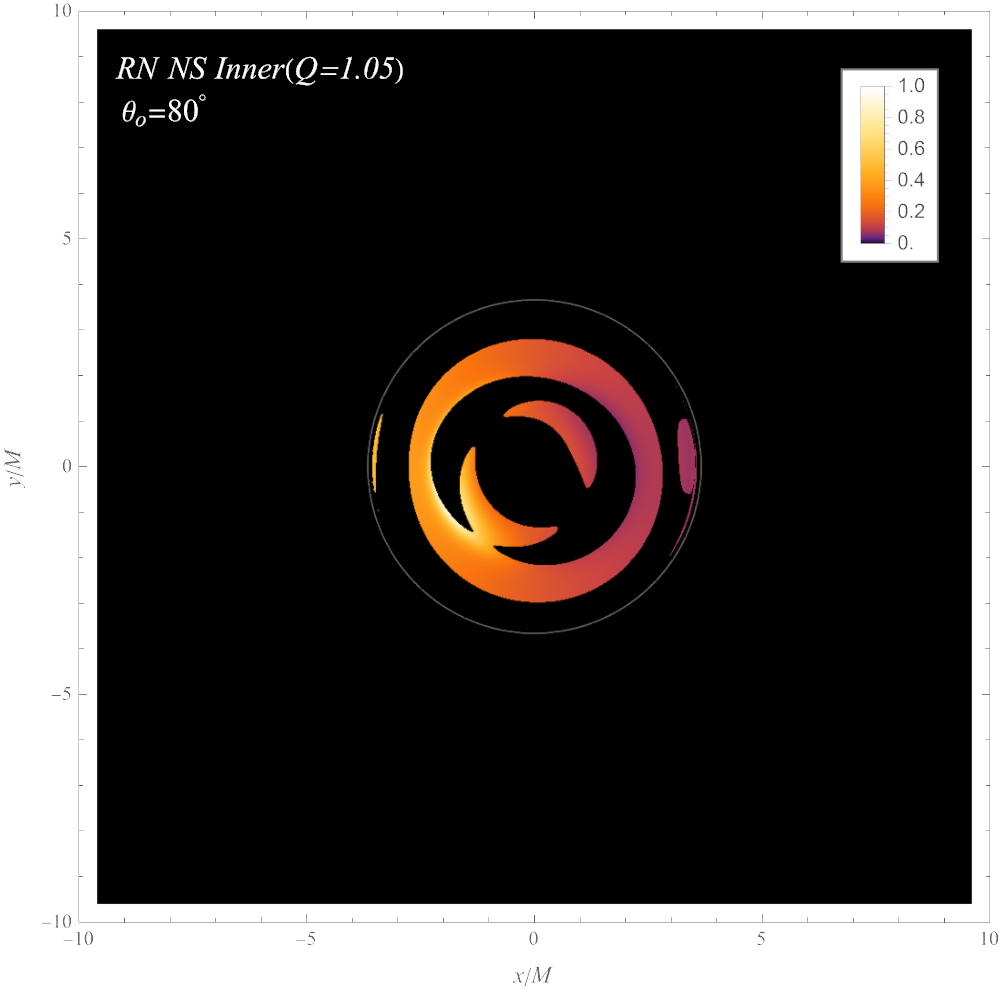}

\includegraphics[width=0.25\textwidth]{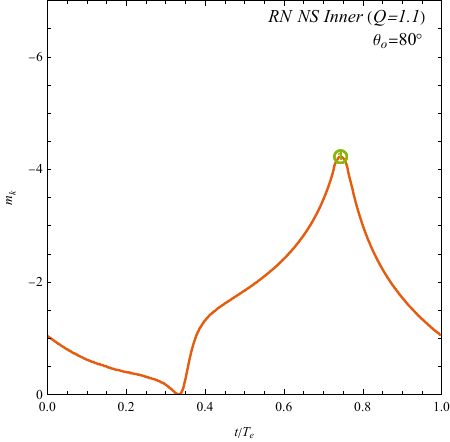}\includegraphics[width=0.25\textwidth]{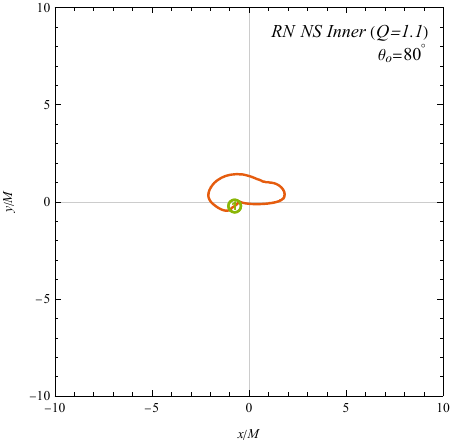}\includegraphics[width=0.25\textwidth]{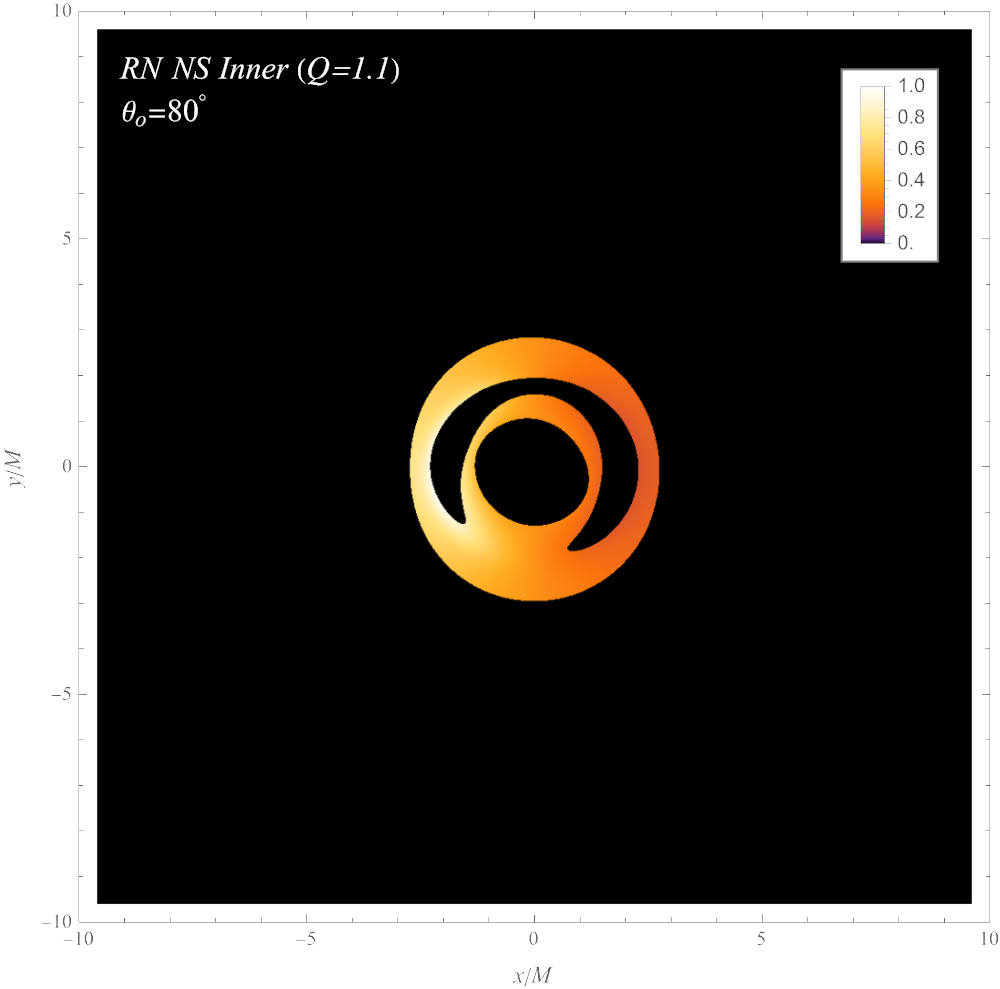}\hfill{}

\caption{Temporal magnitudes $m_{k}$ (\textbf{First Column}) and centroids
$c_{k}$ (\textbf{Second Column}) as a function of $t/T_{e}$ for
cases $Q/M=1.05$ (\textbf{Top Row}) and $Q/M=1.1$ with the orbit
near $r^{*}$ (\textbf{Bottom Row}). The inclination is $\theta_{o}=80^{\circ}$
as well. The highest and second-highest peaks of the temporal magnitude
are indicated by \textcolor{darkpastelgreen}{\textcircled{1}}
and \textcolor{denim}{\textcircled{2}}, respectively. The centroids
of the flux at these peaks are identified with corresponding numbers.
We also provide snapshots for the each cases, when the temporal magnitude
reaches its highest peak (\textbf{Third Column}) and second-highest
peak (\textbf{Fourth Column}). Note that only a single peak appears
in the case of $Q/M=1.1$.}
\label{fig:snapshots new2}
\end{figure}

To achieve a more profound comprehension of the alterations in integrated
images, this part studys the temporal magnitudes $m_{k}$ and centroids
$c_{k}$ observed at an inclination angle of $\theta_{o}=80^{\circ}$
for each cases. FIGs. \ref{fig:snapshots new1} and \ref{fig:snapshots new2}
highlight two significant peak in the temporal magnitude for each
cases, labeled as \textcolor{darkpastelgreen}{\textcircled{1}}
and \textcolor{denim}{\textcircled{2}}. Additionally, corresponding
snapshot images at these specific moments are provided for further
examination.

FIG. \ref{fig:snapshots new1} presents the results for three cases:
the hot spot orbiting the RN black hole with $Q/M=0.9$ at $r_{\text{msco}}$
(top row), the hot spot orbiting the RN singularity with $Q/M=1.05$
at $r_{\text{msco}}$ (middle row), and the hot spot orbiting the
RN singularity with $Q/M=1.1$ at $r_{\text{msco}}^{+}$ (bottom row).
In all cases, the primary image ($n=0^{>}$) generated by the hot
spot close to the leftmost portion of its orbit dominates the flux
at the highest peak (as shown in third column). This aligns with our
expectations based on the Doppler effect, as the hot spot approaching
the observer on the left side experiences a blue shift, leading to
a noticeable increase in the observed flux.

Conversely, as the hot spot moves away from the observer, the primary
image undergoes a phase of diminished flux, reducing its dominant
contribution and allowing higher-order images to generate localized
flux peaks. Notably, the secondary peak in each scenario reveals distinct
contributors:
\begin{itemize}
\item In the $Q/M=0.9$ case, the $n=1^{>}$ image becomes a significant
contributor to the total flux, leading to the secondary peak. 
\item In the $Q/M=1.05$ case, the $n=2^{<}$, $n=3^{<}$ and higher-order
images within critical curve collectively contribute to the secondary
peak. 
\item In the $Q/M=1.1$ case, the absence of a critical curve allows the
$n=2^{<}$ and $2^{>}$ images to emerge as dominant contributors
at the secondary peak. 
\end{itemize}
These distinct contributions to the secondary peak in each cases provide
valuable features for differentiating between them.

The corresponding centroids are presented in the second column of
FIG. \ref{fig:snapshots new1}. When the primary image exhibits relative
dominance, the temporal centroid coincides with the primary image's
center. Doppler-induced flux reduction in the primary image, coupled
with the presence of higher-order images, can substantially displace
the centroid from the center of the primary image's orbit, resulting
in the observed intricate trajectory in the vicinity of the center.
In the case of $Q/M=1.05$, the presence of inner images further accentuates
this intricacy of the centroid. Additionally, for $Q/M=1.1$, the
finite number of higher-order images leads to a smoother centroid.

FIG. \ref{fig:snapshots new2} shows the results for two cases: the
hot spots orbiting the RN black hole with $Q/M=1.05$ (top row) and
$Q/M=1.1$ (bottom row) at $r=1.11M$ and $1.22M$, respectively.
Due to the small radius of the orbit, the hot spot appears proximate
to the observer-singularity line. Notably, the structural symmetry
facilitates the formation of a circular ring in the snapshot image.
Furthermore, the high imaging symmetry contributes to a centroid position
situated closer to the center.

\begin{figure}
\includegraphics[width=0.199\textwidth]{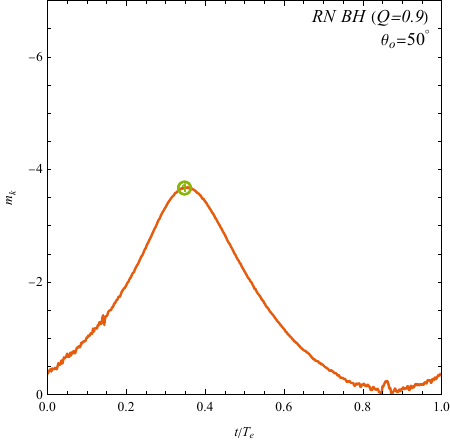}\includegraphics[width=0.199\textwidth]{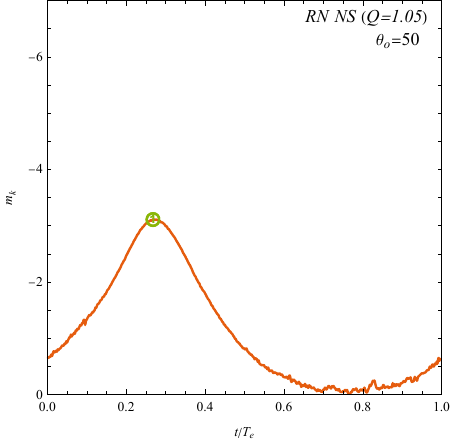}\includegraphics[width=0.199\textwidth]{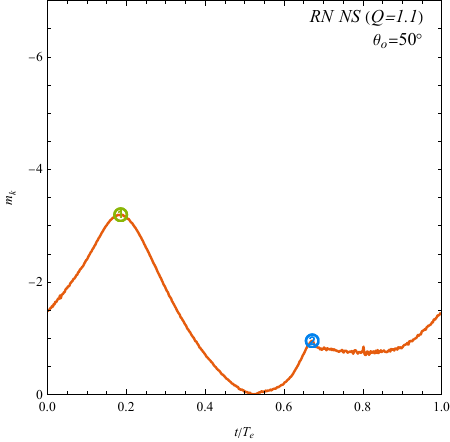}\includegraphics[width=0.199\textwidth]{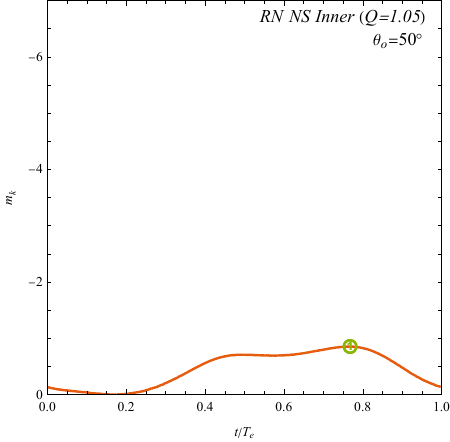}\includegraphics[width=0.199\textwidth]{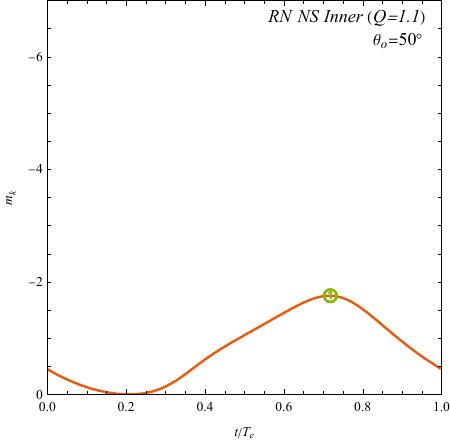}

\includegraphics[width=0.199\textwidth]{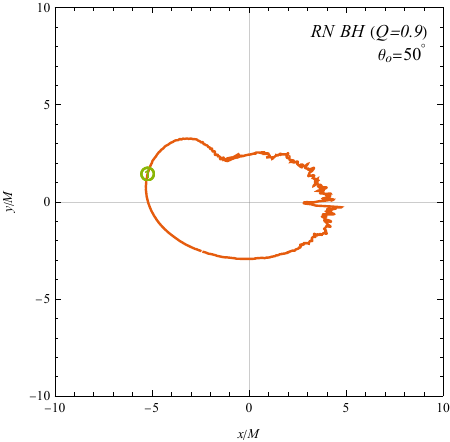}\includegraphics[width=0.199\textwidth]{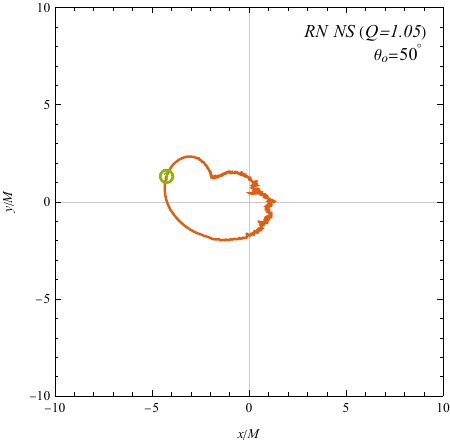}\includegraphics[width=0.199\textwidth]{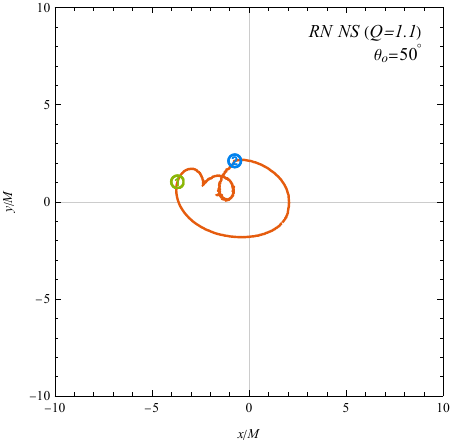}\includegraphics[width=0.199\textwidth]{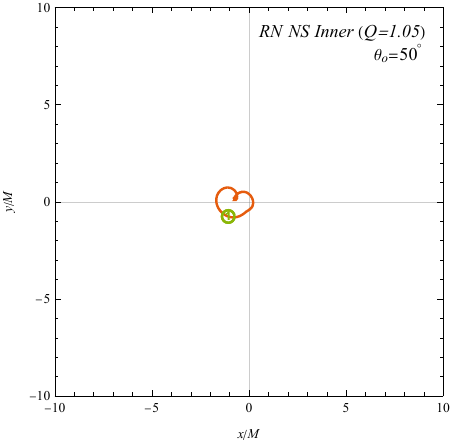}\includegraphics[width=0.199\textwidth]{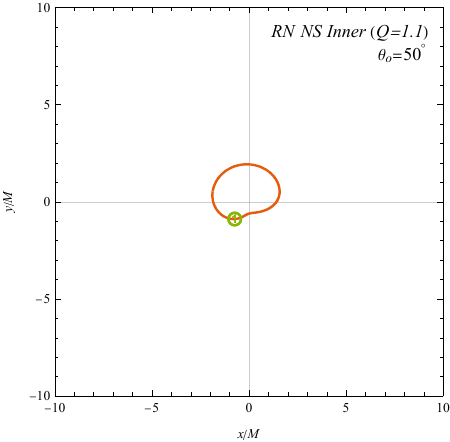}

\caption{Temporal magnitudes $m_{k}$ (\textbf{Upper Row}) and centroids $c_{k}$
(\textbf{Lower Row}) as a function of $t/T_{e}$ for each cases. The
inclination is $\theta_{o}=50^{\circ}$. }
\label{fig:Angle 50 new}
\end{figure}

In the end, we present the temporal magnitudes and centroids for an
inclination angle of $\theta_{o}=50^{\circ}$ in FIG. \ref{fig:Angle 50 new}.
In contrast to the inclination of $\theta_{o}=80^{\circ}$, the temporal
magnitudes at $\theta_{o}=50^{\circ}$ exhibit a single peak. This
difference arises from the reduced influence of the Doppler effect
at the lower inclination angle. As a consequence, the flux becomes
less frequency-dependent, allowing the primary image to dominate the
total flux for a larger portion of the observation period. This, in
turn, minimizes the impact of higher-order images on the centroids,
leading to a smoother and less intricate trajectory.

\section{Conclusions}

\label{sec:CONCLUSIONS} 

This study investigates the observational signatures of RN black holes
and singularities, focusing on the hot spots orbiting celestial bodies.
Interestingly, the presence or absence of a photon sphere, which significantly
impacts the observed features of the orbiting hot spots, surrounding
a RN singularity depends on the charge parameter $Q/M$. To disentangle
the diverse manifestations arising from different scenarios within
RN spacetime, we compare the observational signatures of RN black
holes, RN singularities with a photon sphere, and RN singularities
without photon spheres. The main discrepancy is expected to be more
pronounced at an observational inclination angle of $\theta_{o}=80^{\circ}$,
and we summarize these differences as follows:
\begin{itemize}
\item For the RN black hole case, the observed features closely resemble
those of Schwarzschild black holes: two distinct lensing image tracks
emerge in time-integrated images capturing a complete hot spot orbit.
\item In contrast, for the RN naked singularity with a photon sphere, images
outside the critical curve mirror those of black holes. However, unlike
black holes, RN naked singularities can exhibit additional images
within the critical curve, leading to a distinct second peak in the
magnitude profiles.
\item Intriguingly, the RN naked singularity without photon spheres display
a finite number of tracks in the time-integrated image. Notably, incomplete
and converging tracks appear. This merging and decomposition of tracks
results in a rapid rise and fall of magnitudes, forming a pronounced
second peak.
\end{itemize}
The analysis of these image characteristics serves as a powerful aid
in distinguishing between RN black holes and singularities, and also
contributes to exploring the nature of naked singularities and further
assessing their potential as alternatives to black holes. With the
advent of the next-generation Very Long Baseline Interferometry (VLBI),
the ability to observe finer features is significantly enhanced. This
will undoubtedly opens up a wider range of potential avenues for detecting
the aforementioned characteristics.
\begin{acknowledgments}
We are grateful to Guangzhou Guo and Xin Jiang for useful discussions
and valuable comments. This work is supported in part by NSFC (Grant
No. 12105191, 11947225 and 11875196).
\end{acknowledgments}

\bibliographystyle{unsrturl}
\bibliography{ref}

\end{document}